\documentclass{IEEEtran}
\IEEEoverridecommandlockouts

\ifCLASSINFOpdf
\else
\fi

\usepackage{graphicx}
\usepackage{caption}
\usepackage{subcaption}
\usepackage{latexsym}
\usepackage[utf8x]{inputenc}
\makeatletter
\newlength \figwidth
\if@twocolumn
\else
\fi
\makeatother
\makeatletter
\newlength \figwidthuna
\makeatother
\usepackage{amssymb}
\usepackage{amsfonts}
\usepackage{amsmath}
\usepackage{bm}

\newcommand{\m}[1]{\boldsymbol{\mathbf{#1}}}
\newcommand{\hatm}[1]{\hat{\boldsymbol{\mathbf{#1}}}}

\usepackage{xcolor}
\usepackage{algorithm}
\usepackage{algorithmic}
\usepackage{cite}
\usepackage{url}
 \usepackage{epsfig}
\newcommand{\qed}{\nobreak \ifvmode \relax \else
      \ifdim\lastskip<1.5em \hskip-\lastskip
      \hskip1.5em plus0em minus0.5em \fi \nobreak
      \vrule height0.75em width0.5em depth0.25em\fi}

\newcounter{MYtempeqncnt}

\newcommand{\diag}{\mbox{diag}}

\newcommand{\T}{{\textrm{T}} }
\renewcommand{\vec}{\mbox{vec } }
\newcommand{\tr}{\mbox{tr}}
\newcommand{\field}[1]{\mathbb{#1}}
\begin{document}

%
\title{Learning-Based Adaptive Transmission  for \\ Limited Feedback Multiuser MIMO-OFDM }

\author{Alberto~Rico-Alvariño,~\IEEEmembership{Student~Member,~IEEE} and Robert~W.~Heath~Jr.~\IEEEmembership{Fellow,~IEEE}
\thanks{
 A. Rico-Alvariño is with the Signal Theory and Communications Department, University of Vigo, Galicia, Spain. (e-mail: alberto@gts.uvigo.es).

R. W. Heath Jr. is with the Department of Electrical and Computer Engineering, The University of Texas at Austin, TX, USA. (e-mail: rheath@ece.utexas.edu).

This work was performed while the first author was visiting the Wireless Networking and Communications Group at The University of Texas at Austin. Work by A. Rico-Alvariño supported by a \textit{Fundación Pedro Barrié de la Maza} graduate scholarship. Work by R. W. Heath Jr. supported by the National Science Foundation Grant 1218338 and by the Army Research Laboratory contract W911NF-10-1-0420.

A preliminary version of this work was presented in EUSIPCO 2013 \cite{RicoHeathEusipco13}.
}}

%



\maketitle

\begin{abstract}

Performing link adaptation in a multiantenna and multiuser system is challenging because of the coupling between precoding, user selection, spatial mode selection and use of limited feedback about the channel. The problem is exacerbated by the difficulty of selecting the proper modulation and coding scheme when using orthogonal frequency division multiplexing (OFDM). This paper presents a data-driven approach to link adaptation for multiuser multiple input mulitple output (MIMO) OFDM systems. A machine learning classifier is used to select the modulation and coding scheme, taking as input the SNR values in the different subcarriers and spatial streams. A new approximation is developed to estimate the unknown interuser interference due to the use of limited feedback. This approximation allows to obtain SNR information at the transmitter with a minimum communication overhead. A greedy algorithm is used to perform spatial mode and user selection with affordable complexity, without resorting to an exhaustive search. The proposed adaptation is studied in the context of the IEEE 802.11ac standard, and is shown to schedule users and adjust the transmission parameters to the channel conditions as well as to the rate of the feedback channel.

\end{abstract}
%
%

\section{Introduction}
Dynamically adapting the transmitter in response to changing channel conditions is key to achieving both throughput and reliability on wireless communication links. 
Reconfigurating the link requires adjusting several transmission parameters: the modulation and coding scheme (MCS), the  multiple input multiple output (MIMO) precoding matrices in multiple antenna systems, the spatial mode (number of spatial data streams for each user), and the assignment of transmit resources for the different users, among other parameters. 
These values are selected based on the available channel state information (CSI) at the transmitter (CSIT). Most modern communication systems, including cellular technologies (3GPP LTE \cite{LTE}, IEEE 802.16 \cite{WIMAX}), wireless local area networks (IEEE 802.11 \cite{11-2012}) and satellite standards (DVB-S2 \cite{DVB-S2}, DVB-RCS \cite{DVB-RCS}) include some kind of link adaptation mechanism.

Classic work on link adaptation focused on narrowband fading. For example, the modulation and power were dynamically adjusted with constraints on the uncoded bit error rate (BER) and average transmit power to maximize the spectral efficiency \cite{Goldsmith97, Chung96}. 
The extension of these approaches to coded transmission resulted in more complicated analytical expressions, and usually required the use of BER approximations \cite{Goldsmith98,Liu04}. The MCS selection problem in \cite{Goldsmith97, Chung96,Goldsmith98,Liu04} is essentially a unidimensional problem that consists of assigning a signal to noise ratio (SNR) interval to each MCS. A different approach to link adaptation can be seen in works like \cite{Wang2013,Kamerman97}, where the transmit rate is modified without taking into account the current channel state, but only with ACK/NAK information.

Link adaptation in systems with multiple channels is more challenging due to the higher dimensionality of the CSI. The channel state cannot be characterized with a single SNR value in systems using orthogonal frequency division multiplexing (OFDM) or MIMO with spatial multiplexing. The reason is that different symbols in the same codeword experience different SNR values. Because of the complicated mapping between codeword error and symbol error, the average SNR may not contain enough information to permit effective adaptation \cite{Lampe02}. 
An alternative is to map the set of SNR values (one for each carrier and spatial stream) to one {effective SNR} \cite{Brueninghaus05, Blankenship04, newcom, Lamarca05, Jensen08}. 
The {effective SNR} is defined as the SNR for an additive white Gaussian noise (AWGN) channel to experience the same frame error rate (FER) as the fading channel under study.
{Effective SNR} metrics are defined as a Kolmogorov mean \cite{kolmogorov30} of the SNR values with some parameters that are fitted according to empirical results. The WiMAX forum, for example, recommends the exponential effective SNR metric as the default method for FER prediction \cite{Jain08} in IEEE 802.16e. The effective SNR metrics lead to adaptation algorithms in the form of look-up tables, where each effective SNR value is associated to an MCS. Some works like  \cite{Martorell11} make use of effective SNR metrics to develop link adaptation algorithms in single user scenarios. Having a fixed mapping between effective SNR and MCS is not ideal due to the impact of practical impairments like non-linearities, non-Gaussian noise or implementation-dependent parameters, like Viterbi truncation depth or number of rounds in a turbo decoder.

Data-driven approaches provide a solution to the problem of mapping an appropriate FER to the set of SNR values.
Based on empirical observations of the SNR values and their associated FER, learning algorithms have been designed to select the proper MCS for each channel realization  \cite{DanielsConv, Puljiz11, DanielsNonUniform, DanielsOnlineSVM, DanielsOnline,Wahls13}. 
This {classification} task is performed by machine learning algorithms like $K$-nearest neighbors or support vector machines (SVM) \cite{elementsStat}. These algorithms are usually described as {non-parametric}, since they do not assume any model that maps SNR values to FER, but try to learn it from empirical data.
Learning can be performed online \cite{DanielsOnline,DanielsOnlineSVM,Wahls13} or offline \cite{DanielsNonUniform, DanielsConv}, and is usually based on a low dimensional feature set containing a subset of the ordered SNR values \cite{DanielsOrdered}. 
This data-driven formulation was shown to outperform the {effective SNR} in \cite{DanielsConv}, and is resiliant to practical impairments like non-Gaussian noise or amplifier non-linearities \cite{DanielsOnline}.

Prior work in learning-based link adaptation focused in single user scenarios \cite{DanielsConv, Puljiz11, DanielsNonUniform, DanielsOnlineSVM, DanielsOnline}. The extension of these approaches to the more general case of having multiple users served at the same time by the use of space division multiplexing (SDM) is not trivial due to the interaction between user selection, mode selection, precoding, and MCS selection. For example, the link adaptation algorithm in \cite{DanielsConv} requires running the classification algorithm for all possible number of spatial streams (NSS) and selecting the NSS leading to a higher throughput. Applying a similar strategy in the multiuser case would require an additional exhaustive search over the choice of users and the number of spatial streams per user.

Prior work in learning-based link adaptation also did not consider the impact of limited feedback precoding  \cite{DanielsConv, Puljiz11, DanielsNonUniform, DanielsOnlineSVM, DanielsOnline}. In multiuser MIMO (MU-MIMO) systems, limited feedback creates quantization error that results in residual interuser interference  \cite{Jindal06}. The resulting interference makes performance a function of the feedback quality. Therefore a smart multiuser link adaptation algorithm should predict the interuser interference and move from aggressive multiuser transmission to more conservative modes depending on the feedback quality \cite{Zhang09}.

In this paper, we present a data-driven link adaptation method that includes user selection, mode selection, precoding, MCS selection, and limited feedback. The main objective of the paper is to give insight into the effect of limited feedback in MU-MIMO systems with link adaptation. To do so, our solution assumes perfect CSI at the receivers, and different degrees of CSI at the transmitter. With this information, and taking into account the degradation due to CSI inaccuracy, the transmitter is able to perform link adaptation. Our work is different from previous approaches in several ways. The focus of \cite{Esslaoui11,Esslaoui12} is to study fairness in a multiuser setting, with limited feedback information not taken into account, and MCS selection is performed by means of effective SNR. In \cite{Chen10}, only a single spatial stream is allowed per user, and the effective SNR approach is followed for MCS selection. In \cite{Yun11}, adaptation is performed following a data-driven approach, and limited feedback is taken into account, but the communication scenario is multicast (i.e., there is only a common message for all receivers). Our paper studies the broadcast scenario, takes into account limited feedback, includes the possibility of transmitting more than one spatial stream to each user, and performs MCS selection following a data driven approach. 

We focus on the multiuser capabilities of IEEE 802.11ac \cite{11ac} to develop our link adaptation method. We consider this standard due to its novelty and the challenging constraint of not allowing user allocation among subcarriers, like in LTE or WIMAX. The adaptation is performed at the access point (AP), thus requiring a minimum communication overhead with the user stations (STA) to obtain CSI. This AP-centric approach is specially suitable for link adaptation in scenarios with cheap STAs, as they might not implement some of the optional features in IEEE 802.11ac to perform MCS selection.

The main contributions of the paper are summarized as follows.
\begin{itemize}
\item We derive a closed form approximation of the interuser interference leakage due to the use of block diagonalization precoding with zero forcing (ZF) receivers. This approximation, unlike previous approaches, is not based on a random vector quantization analysis. We exploit the particular codebook structure used in IEEE 802.11ac, which is induced by the use of Givens decompositions, to derive our approximation. This approximation is shown to be very accurate for different feedback rates. We restrict our analysis to ZF receivers as a \textit{worst-case} scenario, so our interference estimation will be conservative if the receiver employs more advanced algorithms, such as minimum mean squared error (MMSE) or maximum likelihood detectors. 
\item We apply previous data-driven approaches\cite{DanielsConv, Puljiz11, DanielsNonUniform, DanielsOnlineSVM, DanielsOnline}, which are limited to the single user, to a multiuser setting with a variable number of spatial streams per user. We show that the MCS selection accuracy of the data-driven approach outperforms unidimensional metrics such as average SNR or effective SNR. The machine learning classifier is shown to be robust to changes in the statistical distribution of the channel, i.e., is able to correctly perform MCS selection even when the training is done with channel samples taken from a different statistical distribution.
\item We use a \textit{greedy} algorithm that performs mode and user selection, inspired on previous work like \cite{Dimic05, Kobayashi06, Chen08}. This algorithm exploits information on the feedback rate, the SNR regime and the number of users to perform user and mode selection. The greedy algorithm sequentially adds spatial layers from the different users until the maximum number of spatial streams is reached, or the throughput is no longer increased. This algorithm, with a complexity that is linear in the maximum number of spatial streams and in the number of users, allows solving the user and mode selection problem without resorting to an exhaustive search.
\end{itemize}

The remaining of the paper is organized as follows: Section \ref{s:ac} describes the mechanisms in IEEE 802.11ac that enable MU-MIMO operation; Section \ref{s:model} presents the system model; Section \ref{s:prEst} introduces the problem statement. The following sections present the main contributions of the paper: Section \ref{ss:precoding} describes the MU-MIMO precoding problem, and presents an approximation for the interference leakage due to limited feedback precoding; Section \ref{s:MCS} describes the data-driven MCS selection; Section \ref{s:userSel} presents a greedy algorithm for user and mode selection; Section \ref{s:results} presents the simulation results; Section \ref{s:conc} concludes the paper.

\textit{Notation:} Matrices are denoted by capital bold letters $\m A, \m B$ and column vectors by lower case bold letters $\m a, \m b$. $\m A^*$ denotes the Hermitian transpose of matrix $\m A$, $\m A^T$ denotes the transpose of matrix $\m A$, $[\m A]_{ij}$ denotes the element in the $i$-th row and $j$-th column of matrix $\m A$, $\m I_K$ denotes the $K\times K$ identity matrix, $\m 0_{K \times N}$ denotes the zero matrix of size $K \times N$, $\m A \circ \m B$ denotes the (entrywise) Hadamard product between matrices $\m A$ and $\m B$,  $| {\cal A} |$ denotes the number of elements in set ${\cal A}$, $\field{C}^{K}$ denotes the set of column vectors with $K$ complex entries, and $\field{C}^{K \times N}$ denotes the set of $K \times N$ matrices with complex entries. ${\cal A} \setminus {\cal B}$ denotes the set difference operation, i.e., $x \in \left({\cal A} \setminus {\cal B} \right) \iff x \in {\cal A} ,\, x \notin {\cal B}$

%
%

%
%
\section{MU-MIMO in IEEE 802.11ac}
\label{s:ac}
IEEE 802.11ac is an emerging wireless standard that supports MU-MIMO, single user MIMO, OFDM modulation, and thus needs sophisticated link adaptation algorithms. In this section we summarize some of the mechanisms that enable MU-MIMO link adaptation in IEEE 802.11ac, and explain how they relate to our system assumptions. Given the functionality required in this paper, we divide the MU-MIMO operation into three different tasks: \textit{CSI acquisition}, \textit{link adaptation} and \textit{MU-MIMO transmission}.
\subsection{CSI acquisition}
Despite being a time division duplexing system, channel reciprocity is not natively supported by IEEE 802.11ac to obtain CSIT.  IEEE 802.11n \cite{11-2012} describes a procedure to obtain CSI at the AP by the transmission of a training sequence in the uplink \cite[Sec. 9.29.2.2]{11-2012} and a calibration procedure to identify the differences between the transmit and receive chains at the STA. These mechanisms are not present in IEEE 802.11ac, so acquisition of CSI based on channel reciprocity cannot rely on any information exchange with the receivers. Although there is some research in estimating the channel from the normal packet exchange in IEEE 802.11ac \cite[Sec. 2.3.3]{11accisco}, we only consider CSI acquisition at the AP by the mechanism described in the standard. This mechanism comprises sending a sounding sequence in the downlink and feeding back the estimated channel to the AP. In the following, we describe these two tasks.

\subsubsection{Sounding}
Channel sounding is initiated by an AP by transmitting a very high throughput (VHT - name of the new transmission modes in IEEE 802.11ac) null data packet (NDP) announcement, which is a control frame that includes the identifiers of the set of users which are potentially going to be polled for feedback. Together with the identifiers, the frame carries information on the requested feedback (there are two different types, single user - SU or multiuser - MU) and the number of columns (spatial streams, only for the MU case) in the requested feedback matrix. This frame is described in more detail in \cite[Sec. 8.3.1.19]{11ac}.

\begin{figure}
\centering
 \includegraphics[width=\figwidth]{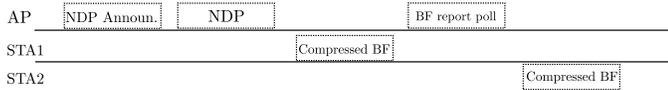}
\caption{Typical message interchange for sounding and feedback.}
\label{f:feedback}
\end{figure}

After the NDP announcement, the AP transmits an NDP, which is used by the receivers to estimate the MIMO channel. After estimating the channel, the first user in the list of the NDP announcement sends feedback information, and the remaining users (if any) transmit their CSI by responding to subsequent beamforming (BF) report polls. This operation is depicted in Figure \ref{f:feedback}.

\subsubsection{Quantization and Feedback}
\label{sss:quant}
The CSI obtained after the training phase is quantized prior to the transmission on the feedback channel. 
The most relevant parameters for the MU-MIMO operation are the preferred beamforming matrices, which contain the right eigenvectors associated to the largest singular values of the channel matrix. 
These matrices are represented using a Givens decomposition. The angles resulting from this decomposition are the parameters that are fed back to the AP.
The feedback message contains the beamforming matrices plus some additional information. This information depends on the selected feedback mode (MU or SU) indicated by the AP in the feedback request message \cite[Sec. 8.4.1.48]{11ac}.

\begin{itemize}
\item {\bf Grouping}: Determines if the carriers are grouped in the feedback message, or if feedback is provided for every carrier. In this paper we assume no grouping.
\item {\bf Codebook information}: Determines the number of bits for the codebook entries. This parameter is selected by the STA. In this paper, we treat different feedback rates, but the problem of selecting an appropriate feedback rate is out of the scope of this work.
\item {\bf Beamforming matrices and SNR information}: Includes preferred beamforming matrix, average SNR for each spatial stream, and SNR for each carrier (or group of carriers) if feedback is in MU mode.
\end{itemize}

\subsection{Link adaptation}
The objective of link adaptation is to perform user selection, mode selection, MCS selection and MIMO precoding design. This can be done in a centralized way at the transmitter, taking as input the limited feedback described in the previous section, or in a distributed way by requesting an STA to select the preferred mode and MCS for a given MIMO configuration.
In this paper, we follow a centralized approach, since our solution involves joint scheduling and MCS selection. 
%

\subsection{MU-MIMO transmission}
Once the users, MCS, mode and precoders are selected, MU-MIMO transmission takes place. MU-MIMO in IEEE 802.11ac is \textit{non-transparent}, meaning that the STAs are aware they will be jointly scheduled with other STAs. This allows a given STA to use the training sequences not only to estimate its own MIMO channel, but also the interference \cite[Sec. 22.3.11.4]{11ac}, which allows performing some advanced tasks such as the modification of the receive precoders, or the correct calculation of the Viterbi weights for decoding. 

%
%

\section{System Model}
\label{s:model}

Consider the downlink of an $N$-carrier OFDM wireless network where an AP equipped with $N_\mathrm{tx}$ antennas communicates with $U$ STAs, ${\cal U} = \left\{ 1,\, \ldots,\, U\right\}$ where the $u$-th user has $N_{\mathrm{rx},u}$ receive antennas. At a given time instant, the transmitter conveys information to a subset of the users ${\cal T} \subseteq {\cal U}$. In a given time slot all users use all subcarriers. 

We assume that the transmitter employs linear precoding, and the receivers use linear equalizers. A single modulation ${\cal M}_u = \left\{m_1 \ldots m_{|{\cal M}_u|} \right\} \subset \field{C}$ is selected for the $u$-th user, constant over all carriers and spatial streams\footnote{Although in IEEE 802.11n the use of different modulations in each spatial stream was allowed, it was apparently not implemented in most commercial devices, and finally discarded for 802.11ac.}.  For a given carrier $n$, $\m s_u[n] \in {\cal M}_u^{L_u}$  is the $L_u$ spatial streams modulated signal containing the information for the $u$-th user, $\m F_u[n] \in \field{C}^{N_\mathrm{tx} \times L_u}$ is the transmit precoding matrix for the $u$-th user, $\m H_u[n] \in\field{C}^{ N_{\mathrm{rx},u} \times N_\mathrm{tx} }$ is the flat-fading MIMO channel from the transmitter to the $u$-th receiver, $\m B_u[n] \in \field{C}^{L_u \times N_{\mathrm{rx},u}}$ is  the interference removal matrix, and $\m G_u[n] \in \field{C}^{ L_u \times L_u}$ is the linear equalizer applied at the $u$-th receiver. We divide the receive processing into two different matrices for simplicity in the treatment of the multiuser precoding. The objective of $\m B_u[n]$ is to reject the interuser interference, while the equalizer $\m G_u[n]$ removes the intrauser interference. To limit the transmit power per carrier, we define the power normalization factor 
$
P[n] \triangleq \sum_{u\in{\cal T}} \tr\left(\m F_u[n]\m F_u^*[n] \right).
$
We use $\m n_u[n] \sim {\cal CN} \left( \m 0,\sigma^2\m I\right)$ to denote the received noise vector at the $u$-th receiver. Given these definitions, the post-processed signal at the $u$-th receiver $\m y_u[n] \in \field{C}^{L_u}$ is
\begin{eqnarray}
 \m y_u[n] &=&  \m G_u[n] \m B_u[n] \m H_u[n] \sum_{i\in{\cal T}}\frac{1}{\sqrt {P[n]}} \m F_i [n] \m s_i[n] \\ \nonumber &&+  \m G_u[n] \m B_u[n] \m n_u[n]
\end{eqnarray}
with $\field{E}\left(\m s_u[n] \m s_u^*[n] \right) = \m I_{L_u}$. For the sake of clarity, we define 
$\hat {\m H}_{u,i}[n]\triangleq \frac{1}{\sqrt {P[n]}} \m G_u[n] \m B_u[n] \m H_u[n] \m F_i [n],$ and $\m w_u[n] \triangleq \m G_u[n] \m B_u[n] \m n_u[n]$, so
\begin{equation}
\label{e:interUserInt}
 \m y_u[n] = \hat {\m H}_{u,u}[n] \m s_u[n] + \underbrace{\sum_{i\in{\cal T},\, i \neq u } \hat {\m H}_{u,i}[n] \m s_i[n]}_{\mbox{Interuser interference}} + \underbrace{\vphantom{\sum_{i\in{\cal T},\, i \neq u }}\m w_u[n]}_{\mbox{Noise}}.
\end{equation}

The transmit signal for each of the scheduled receivers is the result of performing coding, interleaving and constellation mapping operations on a stream of source bits. The MCS for the $u$-th user $c_u$ is selected from a finite set of MCS ${\cal C}$. The selected number of spatial streams and MCS for the $u$-th user has an associated rate of $\eta\left(c_u, L_u \right)$ bits per second.

In general, the probability that a frame is not correctly decoded at the $u$-th receiver (i.e., the FER), depends on the transmit power, channel matrices, number of scheduled users, selected MCS for the $u$-th user, and selected modulation for the interfering users. By treating the noise and residual multi-user interference as Gaussian, and assuming a linear receiver, it is reasonable to write the FER of the $u$-th user $p_u$ as a function of the the selected MCS $c_u$ and the post-processing SNR values $\displaystyle\m \gamma_u = \left[ \gamma_{u,1}\left[1\right], \, \ldots,\, \gamma_{u,1}\left[N\right],\,\ldots,\, \gamma_{u,L_u}\left[1\right],\, \ldots,\, \gamma_{u,L_u}\left[N\right] \right]^T$, where
\begin{equation}
\label{e:FER}
 p_u = \mathrm{FER }\left(\m \gamma_u, {c}_u \right).
\end{equation}
The post processing SNR of the $u$-th user in the $i$-th spatial stream and $n$-th carrier is defined as
\begin{equation}
\label{e:postSNR}
 \gamma_{u,i}[n] = \frac{\left|\left[ \m D_{u,u}\left[n\right]\right]_{ii}\right|^2}{\left[ \m R_{u}[n] \right]_{ii}}\,\,\, u = 1\ldots U, i=1\ldots L_u
\end{equation}
with
\begin{eqnarray}
\label{e:covMatrix}
 \m R_{u}[n] = \left( \hat {\bf H}_{u,u}[n] - \m D_{u,u}[n] \right) \left( \hat {\bf H}_{u,u}[n] - \m D_{u,u}[n] \right)^* + \\ \nonumber  \displaystyle\sum_{j \in {\cal T},\, j \neq u} \hat {\bf H}_{u,j}[n]\hat{ \m H}^*_{u,j} [n] + \sigma^2  \m G_u[n] \m B_u[n]\m B_u^*[n]\m G_u^*[n] 
\end{eqnarray}
the covariance matrix of interference plus noise, and $\m D_{u,u} \triangleq  \m{\hat H}_{u,u} \circ \m I_{L_u}$. Based on the rate $\eta\left(c_u,L_u \right)$ and the FER $ p_u$, we  define the throughput of user $u$ as 
\begin{equation}
\label{e:throughput}
 t_u =  \eta\left(c_u,L_u \right) \left(1- p_u \right).
\end{equation}

%
%

%
%
\section{Problem Statement}
\label{s:prEst}
In this section, we formulate the link adaptation problem. The adaptation problem in the multiuser scenario is different from the single user scenario. In the single user case, the usual objective of link adaptation is to maximize the (unique) link throughput subject to a constraint on the FER. In the MU-MIMO case, each user has a different rate, so the objective might be to maximize a function of the rates, subject to a FER constraint $p_0>0$ (assumed to be equal for all receivers). We consider the sum rate as the performance objective in this paper. If we denote by $\m t = \left[t_1\ldots t_U \right]$ the vector containing the throughput (\ref{e:throughput}) of all users, and by $\nu\left( \m t \right) \triangleq \sum_{u=1}^{U}{t_u}$ the sum rate, the LA problem can be stated as
\begin{equation}
\label{e:optEig}
  \begin{array}{l l l l l }
 \mbox{maximize } & \nu\left( \m t \right) & 
 \mbox{subject to} & p_u \leq p_0&\,\, u=1,\,\ldots,\, U.
\end{array}
\end{equation}
We will assume $t_u = 0,\, p_u = 0$ if $u \notin {\cal T}$ to be consistent with our approach, which involves scheduling a subset of the users. Note that the LA problem can be modified to maximize another utility metric besides the sum rate just by defining $\nu\left( \m t \right)$ accordingly. For example, we can include proportional fairness in this setting by changing the objective function to $\nu\left( \m t \right) =   \sum_{u=1}^{U}\log {t_u}$ \cite{Kim04}.

Trying to solve (\ref{e:optEig}) directly is computationally intractable. Besides the difficulty of obtaining a mathematical model that maps the CSI to the FER $p_u$, the number of design variables is quite large and difficult to handle. The design variables include, among others, the set of active users ${\cal T}$, the streams per each active user $L_u$, the precoding matrices $\m F_u[n]$, the interference removal matrices $\m B_u[n]$, equalizers $\m G_u[n]$ and MCS $c_u$. Moreover, imperfect CSI at the transmitter creates unknown interference leakage between the different receivers. We propose a procedure for link adaptation that finds a good solution to the sum rate maximization problem. Our solution has three operational blocks: \textit{precoding and equalization with interference estimation}, \textit{ MCS selection} and \textit{user and mode selection}.

\section{Precoding and Equalization with Interference Estimation}
\label{ss:precoding}
In this section, we present the precoding technique for MU-MIMO transmission and obtain a closed form approximation for the residual interuser interference caused by limited feedback precoding. Given the subset of active users ${\cal T}$ and the number of streams per user $L_u$, the joint problem of selecting optimum precoders\footnote{As the design of precoders is independent for each carrier, we will drop the index $[n]$ in this section.} $\m F_u$, interference rejection matrices $\m B_u$, and equalizers $\m G_u$ to maximize the sum rate does not have a closed form solution. Several approaches have been proposed, e.g. block diagonalization (BD) \cite{Spencer04} or minimizing signal \textit{leakage} \cite{Sadek07}.
 In this paper we assume that the precoders and interference rejection matrices are obtained using BD. We choose this precoding algorithm because of its simplicity and its low gap to capacity when used in conjunction with user selection algorithms \cite{shen06}.
The proposed link adaptation framework, however, can be used with other precoding techniques, although we restrict our analysis to BD.

\subsection{Block diagionalization precoding}
BD precoding removes the interference between the different users but not the interference between streams associated to the same user. It is well suited for IEEE 802.11ac because the precoders can be designed based on the feedback information provided. The BD precoder is designed as follows. Let 
$
 \m H_u = \m U_u \m \Sigma_u \m V^*_u
$ 
be the SVD decomposition of $\m H_u$ with the singular values in $\m \Sigma_u$ arranged in decreasing order. Note that $\m U_u \in \field{C}^{{N_{\textrm{rx}, u}} \times {N_{\textrm{rx}, u}}}$, $\m \Sigma_u \in \field{R}^{{N_{\textrm{rx}, u}} \times N_{\textrm{tx}}}$ and $\m V_u \in \field{C}^{N_{\textrm{tx}} \times N_{\textrm{tx}}}$. The interference rejection matrix $\m B_u$ is formed by taking the first $L_u$ columns of $\m U_u$ (corresponding to the left singular vectors associated to the $L_u$ largest singular values). Let us denote $\tilde {\m H}_u \triangleq \m  B_u \m H_u$, and
\begin{equation}
 \bar {\m H}_u \triangleq \left[\begin{array}{cccccc}\m {\tilde H}_1  & \ldots & \m {\tilde H}_{u-1} & \m {\tilde H}_{u+1} & \ldots  & \m {\tilde H}_{T}\end{array} \right]^T.
\end{equation}
BD requires that the precoder F satisfies
\begin{equation}
\label{e:ZF}
\bar {\m H}_u \m F_u = \m 0\, \forall u.
\end{equation}
The set of precoders achieving (\ref{e:ZF}) can be written as $\m N_u \m P_u$, where $\m N_u$ is a basis of the nullspace of $\bar {\m H}_u$, and $\m P_u$ is an arbitrary matrix, that can be used to select the directions of transmission (in case the nullspace of $\bar {\m H}_u$ is of dimension higher than $L_u$) as well as to perform power allocation. We choose $\m F_u$ as the matrix containing the $L_u$ singular vectors associated to the largest singular values of $\m {\tilde H}_u \m N_u$, i.e., we perform uniform power allocation along the $L_u$ stronger directions of the equivalent channel $\m {\tilde H}_u \m N_u$. 

The post-processed signal at the $u$-th receiver is 
\begin{equation}
\label{e:BDreceived}
 \m y_u =  \m G_u \m B_u \m H_u \m F_u \m s_u + \m w_u.
\end{equation}
Equation (\ref{e:BDreceived}) can be obtained from (\ref{e:interUserInt}) just by noticing that the BD condition (\ref{e:ZF}) is equivalent to $\m B_u \m H_u \m F_i = \m 0\,\, \forall u \neq i$ and therefore, $\hat {\m H}_{u,i}[n] = \m 0,\,\, \forall u\neq i$.
Assuming limited feedback of CSI, as anticipated in Section \ref{s:ac}, full knowledge of $\m H_u$ is not possible. We now explain how BD can be performed also with reduced information. Decompose $\m H_u$ using the SVD as
\begin{equation}
 \m H_u = \left[ \begin{array}{ll} \m {\tilde U}_{u} & \m {\tilde U}_{u,\textrm{small}}\end{array}\right] \left[ \begin{array}{lll}\m \Sigma_{u,0} & \m 0\\ \m 0 &\m \Sigma_{u,1}  \end{array}\right]\left[ \begin{array}{c}\m {\tilde V}_{u}^* \\ \m {\tilde V}_{u,\textrm{small}}^* \end{array}\right].
\end{equation}
where $ \m {\tilde U}_{u} \in \field{C}^{N_{\textrm{rx},u} \times L_u}$, $ \m {\tilde U}_{u,\textrm{small}} \in \field{C}^{N_{\textrm{rx},u} \times (N_{\textrm{rx},u} - L_u)}$ are the matrices containing the left singular vectors associated to the $L_u$ largest singular values and $N_{\textrm{rx},u} - L_u$ smallest singular values, respectively; $\m \Sigma_{u,0} \in \field{C}^{L_u \times L_u}$ is the diagonal matrix containing the $L_u$ largest singular values of $\m H_u$; and $\m {\tilde V}_{u}^* \in \field{C}^{L_u \times N_{\textrm{tx}}}$,  $ \m {\tilde V}_{u,\textrm{small}}^*  \in \field{C}^{(N_{\textrm{rx},u} - L_u) \times N_{\textrm{tx}}}$ are the matrices containing the $L_u$ right singular vectors associated with the $L_u$ largest singular values, and the $(N_{\textrm{rx},u} - L_u)$ right singular vectors associated to the remaining nonzero singular values. Assuming that the receiver uses $\m B_u = \m {\tilde U}_{u}^*$, then the \textit{equivalent channel} can be written as
$
 \m {\tilde H}_{u} = \m \Sigma_{u,0} \m {\tilde V}_u^*.
$
It can be seen that $\m {\tilde H}_u$ can be available at the transmitter (with some quantization error) with the feedback scheme described in Section \ref{s:ac} for IEEE 802.11ac, since the matrix $\m {\tilde V}_u$ is the beamforming matrix for $L_u$ spatial streams, and the values of $\m \Sigma_{u,0}$ can be obtained from the SNR of each subcarrier and spatial stream. Further, since $\m \Sigma_{u,0}$ is invertible, the nullspace of $\m {\tilde H}_u$ is the same as that of  $\m {\tilde V}_u^*$, so the SNR information is not necessary to design the precoders. Consequently, the precoders can be found by ensuring that 
$
 \m {\tilde V}_u^* \m F_j = \m 0\,\,\forall u \neq j.
$
Therefore, the set of precoding matrices can be obtained from the preferred beamformers $\m {\tilde V}_u$. The presence of limited feedback, however, creates some unknown interference leakage between the different users, which has to be estimated. We present next the quantization scheme in IEEE 802.11ac, and an analytical approximation to this interference leakage.

\subsection{Quantization}
The presence of limited feedback creates unknown interference leakage among the different users. 
The post-processing SNR depends on the interference leakage, so this value has to be estimated.
This estimation can be easily performed at the receive side, and CSIT can be acquired by the use of the feedback link. This procedure, however, is not desirable for various reasons. First, there is a circular dependency between user and mode selection and interference leakage estimation. User and mode selection requires knowing the post processing SNR and, consequently, the interference leakage, but estimating the interference leakage is restricted to a certain user/mode configuration. Second, the amount of overhead and training is roughly doubled with respect to the simple message interchange in Figure \ref{f:feedback}. Therefore, it is desirable to estimate the interference leakage at the transmit side, without additional feedback from the receivers. 

Next we derive an approximation for the interference leakage caused by the quantization procedure in IEEE 802.11ac. This approximation can be easily computed at the transmitter by using a statistical characterization of the quantization error. We describe first the quantization method used in IEEE 802.11ac, and then derive an approximation for the interference leakage.

The objective of the quantization task is to provide the transmitter with a quantized version $\hatm V_u$ of matrix $\m {\tilde V}_u$, which is the preferred beamforming matrix. The quantization process proceeds as follows. The unitary matrix $\m {\tilde V} \in \field{C}^{N_{\textrm{tx}},L}$ is decomposed by using the Givens decomposition \cite[Ch. 5]{golub} as
\begin{equation}
 \m {\tilde V} = \left(\prod_{\ell=1}^{L} \m D_\ell\left(\phi_{\ell,1} \ldots \phi_{\ell,N_{\textrm{tx}}-\ell+1}  \right) \prod_{n=\ell+1}^{N_{\textrm{tx}}} \m G_{n,\ell}\left( \psi_{\ell,n} \right)\right) \m {\tilde I}
\end{equation}
where 
\begin{equation}
\label{e:Dl}
 \m D_\ell\left(\phi_{\ell,1} \ldots \phi_{\ell,N_{\textrm{tx}}-\ell+1}  \right) = \diag\left(\m 1_{\ell-1}, e^{j\phi_{\ell,1}} \ldots e^{j\phi_{\ell,N_{\textrm{tx}}-\ell+1}} \right) \in \field{C}^{N_{\textrm{tx}} \times N_{\textrm{tx}}}.
\end{equation}
$ \m {\tilde I}$ is a matrix containing the first $L$ columns of an $N_{\textrm{tx}} \times N_{\textrm{tx}}$ identity matrix and $\m G_{n,\ell}\left( \psi_{\ell,n} \right) \in \field{R}^{N_{\textrm{tx}} \times N_{\textrm{tx}}}$ is a rotation matrix operating in the $\ell$ and $n$ coordinates:
\begin{equation}
 \m G_{n,\ell}\left( \psi_{\ell,n} \right) = \left[ \begin{array}{ccccc}
                                            \m I_{\ell-1} &&&& \\
						& \cos \psi_{\ell,n} & & \sin \psi_{\ell,n} & \\
						&		  & \m I_{n-\ell-1} & & \\
						& -\sin \psi_{\ell,n} & & \cos \psi_{\ell,n} & \\
						&		   & & & \m I_{N-n}
                                           \end{array}\right].
\end{equation}
The feedback consists of quantized versions  $\hat \psi_{\ell,n}$ and $\hat \phi_{\ell,n}$ of the angles $\psi_{\ell,n}$ and $\phi_{\ell,n}$. The quantization of  $\psi_{\ell,n}$ and $\phi_{\ell,n}$ is performed by the use of $b_\psi$ and $b_\phi$ bits, respectively. IEEE 802.11ac uses a uniform quantizer. Since $\psi_{\ell,n} \in \left[0,\pi/2 \right]$ and $\phi_{\ell,n} \in \left[0,2\pi \right]$ \cite{Roh07}, the codebook for each of the angles is
\begin{equation}
 \hat \psi_{\ell,n} \in \left\{ q_{\psi,k} \triangleq \frac{k\pi}{2^{b_\psi+1}} + \frac{\pi}{2^{b_\psi+2}},\,\, k = 0,1\ldots 2^{b_\psi}-1\right\}
\end{equation}
\begin{equation}
 \hat \phi_{\ell,n} \in \left\{ q_{\phi,k} \triangleq \frac{k\pi}{2^{b_\phi-1}} + \frac{\pi}{2^{b_\phi}},\,\, k = 0,1\ldots 2^{b_\phi}-1\right\}.
\end{equation}
The quantization of the angles is performed by finding the minimum distance codeword
\begin{equation}
 \hat \psi_{\ell,n} = q_{\psi,k} \mbox{ if }  \psi_{l,n} \in {\cal Q}_{\psi,k} \triangleq \left[ \frac{k\pi}{2^{b_\psi+1}} , \frac{(k+1)\pi}{2^{b_\psi+1}} \right]
\end{equation}

\begin{equation}
 \hat \phi_{\ell,n} = q_{\phi,k} \mbox{ if }  \phi_{l,n} \in {\cal Q}_{\phi,k} \triangleq  \left[ \frac{k\pi}{2^{b_\phi-1}} , \frac{(k+1)\pi}{2^{b_\phi-1}} \right].
\end{equation}
For the sake of simplicity, we denote 
$
 \delta \triangleq \frac{\pi}{2^{b_\phi}}
$
and
$
 \epsilon \triangleq \frac{\pi}{2^{b_\psi+2}}
$, 
 so that ${\cal Q}_{\psi,k} = \left[q_{\psi,k} - \epsilon, q_{\psi,k} + \epsilon \right]$ and ${\cal Q}_{\phi,k} = \left[q_{\phi,k} - \delta, q_{\phi,k} + \delta \right]$. Note that this quantization scheme is the LBG quantizer \cite{LBG}, optimal in the minimum distortion sense, only if the angles are independent and the distribution of both angles is uniform, which is not the case even in the well-studied independent and identically distributed Gaussian MIMO channel, see e.g. \cite{Ansari06}.

\subsection{Interference estimation}
\label{ss:intEst}
Now we characterize the residual interference using only the quantized CSI. The BD precoders are designed using the quantized beamformers $\hatm V_u$ to satisfy $\hatm V_u^* \m F_j = \m 0\,\,\forall u \neq j$. Because the beamformers are quantized, the interuser interference cannot be completely removed due to the imperfect CSI, so the total interference plus noise covariance matrix (\ref{e:covMatrix}) assuming BD precoding and ZF equalizer at user $u$ is
\begin{equation}
\label{e:Ru}
 \m R_{u} = \displaystyle\sum_{j \in {\cal T} \setminus \{ u \}}\m  R_{u,j} + \sigma^2  \m G_u \m G_u^*
\end{equation}
where $\m R_{u,j} \triangleq \hat {\m H}_{u,j}\hat{ \m H}^*_{u,j}$ is the covariance matrix of the interference from the message intended to user $j$. Equation (\ref{e:Ru}) follows by assuming perfect CSI at the receiver, so that the ZF equalizer removes all the intra-stream interference, i.e., $\hatm H_{u,u} = \m D_{u,u}$ in  (\ref{e:covMatrix}), and by applying the fact that $\m B_u$ is a unitary matrix. If $\m G_u$ is a ZF equalizer, and the precoders and interference rejection matrices are designed following the BD procedure, then
\begin{equation}
\label{e:gu}
\m G_u  = \left( \m B_u \m H_u \m F_u \right)^{-1} = \left( \m \Sigma_{u,0} \m {\tilde V}_{u}^* \m F_u \right)^{-1}.
\end{equation}
Now, write $\m {\tilde V}_{u}^* = \m {\hat V}_{u}^* + \m E_u$, where $\m E_u$ is the quantization error matrix. If the quantization error is small, (\ref{e:gu}) can be approximated as 
\begin{equation}
\label{e:gu2}
\m G_u =  \left( \m \Sigma_{u,0} \m {\hat V}_{u}^* \m F_u + \m \Sigma_{u,0} \m E_u \m F_u \right)^{-1} \approx  \left( \m \Sigma_{u,0} \m {\hat V}_{u}^* \m F_u \right)^{-1}.
\end{equation}
Note that for (\ref{e:gu2}) to hold it is only necessary that the entries of $\m E_u$ are negligible with respect to the entries of $\m {\hat V}_{u}^*$.

The multiuser interference can be written as 
\begin{eqnarray}
\label{e:Ruj}
\m R_{u,j} &=& \frac{1}{P} \m G_u \m \Sigma_{u,0} \m {\tilde V}_u^* \m F_j \m F_j^*  \m {\tilde V}_u \m\Sigma_{u,0}^* \m G_u^*  \\ \nonumber &=& \frac{1}{P} \m G_u \m \Sigma_{u,0} \m {E}_u^* \m F_u \m F_u^*  \m {E}_u \m \Sigma_{u,0}^* \m G_u^*
\end{eqnarray}
where the last equality is due to the BD constraint $\m {\hat V}_u^* \m F_j = \m 0$. 

\begin{figure*}[t]
\setcounter{MYtempeqncnt}{\value{equation}}
\setcounter{equation}{32}

\begin{equation}
\label{e:eWell}
\m W_{\ell,n} = \field{E} \left[ \begin{array}{ccccc}
                                            \m I_{\ell-1} \otimes \m G_{\ell,n}^* &&&& \\
						& \cos \psi_{\ell,n}\m G_{\ell,n}^* & & -\sin \psi_{\ell,n} \m G_{\ell,n}^* & \\
						&		  & \m I_{n-\ell-1} \otimes \m G_{\ell,n}^* & & \\
						& \sin \psi_{\ell,n} \m G_{\ell,n}^* & & \cos \psi_{\ell,n} \m G_{\ell,n}^* & \\
						&		   & & & \m I_{N_{\textrm{tx}}-n} \otimes \m G_{\ell,n}^*
                                           \end{array}\right]
\end{equation}
\hrulefill
\vspace*{4pt}
\end{figure*}
\begin{figure*}[t]
\begin{align}\label{e:firstEq}
\field{E}\m G_{\ell,n}^* &=&&  \left[ \begin{array}{ccccc}
                                            \m I_{\ell-1} &&&& \\
						& \displaystyle\frac{\sin \epsilon\cos \hat \psi_{\ell,n}}{\epsilon}& & - \displaystyle\frac{\sin \epsilon\sin \hat \psi_{\ell,n}}{\epsilon} & \\
						&		  & \m I_{n-\ell-1} & & \\
						&\displaystyle \frac{\sin \epsilon\sin \hat \psi_{\ell,n}}{\epsilon} & &\displaystyle \frac{\sin \epsilon\cos \hat \psi_{\ell,n}}{\epsilon} & \\
						&		   & & & \m I_{N_{\textrm{tx}}-n}
                                           \end{array}\right]&\\
\field{E}\cos \psi_{\ell,n}\m G_{\ell,n}^* &=&&  \left[ \begin{array}{ccc}
                                             \frac{\cos \hat \psi_{\ell,n} \sin \epsilon}{\epsilon} \m I_{\ell-1} &&\\
						& \m J_{\ell,n} & \\
						   & &  \frac{\cos \hat \psi_{\ell,n} \sin \epsilon}{\epsilon} \m I_{N_{\textrm{tx}}-n}
                                           \end{array}\right]&\\
\noalign{with}
\m J _{\ell,n}&=&& \left[
\begin{array}{ccc}
 \frac{\epsilon + \cos \epsilon \cos 2\hat \psi_{\ell,n} \sin \epsilon}{2\epsilon}& & - \frac{\cos \epsilon \cos \hat \psi_{\ell,n} \sin \epsilon \sin \hat \psi_{\ell,n}}{\epsilon}  \\
								  & \frac{\cos \hat \psi_{\ell,n} \sin \epsilon}{\epsilon} \m I_{n-\ell-1} & \\
						\frac{\cos \epsilon \cos \hat \psi_{\ell,n} \sin \epsilon \sin \hat \psi_{\ell,n}}{\epsilon} & &\frac{\epsilon + \cos \epsilon \cos 2\hat \psi_{\ell,n} \sin \epsilon}{2\epsilon}  \\		
\end{array}
\right]\\
\noalign{and}
\label{e:finalExp}
\field{E}\sin \psi_{\ell,n}\m G_{\ell,n}^* &=&&  \left[ \begin{array}{ccc}
                                             \frac{\sin \hat \psi_{\ell,n} \sin \epsilon}{\epsilon} \m I_{\ell-1} && \\
						& \m S_{\ell,n} & \\
						& & \frac{\sin \hat \psi_{\ell,n} \sin \epsilon}{\epsilon} \m I_{N_{\textrm{tx}}-n}
                                           \end{array}\right]\\
\noalign{with}
\m S_{\ell,n} &=&& \left[\begin{array}{ccc}
\frac{\cos \epsilon \cos \hat \psi_{\ell,n} \sin \epsilon \sin \hat \psi_{\ell,n}}{\epsilon}& & - \frac{\epsilon - \cos \epsilon \cos 2\hat \psi_{\ell,n} \sin \epsilon}{2\epsilon}\\
								  & \frac{\sin \hat \psi_{\ell,n} \sin \epsilon}{\epsilon} \m I_{n-\ell-1} &  \\
						 \frac{\epsilon - \cos \epsilon \cos 2\hat \psi_{\ell,n} \sin \epsilon}{2\epsilon}& &\frac{\cos \epsilon \cos \hat \psi_{\ell,n} \sin \epsilon \sin \hat \psi_{\ell,n}}{\epsilon}  
\end{array}\right].&\label{e:lastEq}
\end{align}
\setcounter{equation}{\value{MYtempeqncnt}}
\hrulefill
\vspace*{4pt}
\end{figure*}
First, note that the interference covariance matrices $\m R_{u,j}$ in (\ref{e:Ruj}) are random variables from the transmitter's point-of-view, since the quantization noise $\m E_u$ is unknown at the transmitter. We define a new covariance matrix by averaging over the realizations of $\m E_u$
\begin{equation}
\label{e:approxInt}
\m {{\bar {R}}}_{u,j} \triangleq {\field E}\,  \m R_{u,j} = \frac{1}{P} \m G_u \m \Sigma_{u,0} \m C_{u,j} \m \Sigma_{u,0}^* \m G_u^* 
\end{equation}
with
\begin{equation}
\label{e:Cki}
 \m C_{u,j} = \field{E}\,\m {\tilde V}_u^* \m F_j \m F_j^* \m {\tilde V}_u
\end{equation}
and the expectation is taken over the random realization of $\m {\tilde V}_u$ given the received feedback $\hatm V_{u}$. Note that we are implicitly averaging over $\m E_u$, but the particular structure of the quantization method makes it easier to derive the final result when explicitly averaging over  $\m {\tilde V}_u | \hatm V_{u}$.

Let us define $\m \phi_\ell \triangleq \left(\phi_{\ell,1},\, \ldots,\, \phi_{\ell,N_{\textrm{tx}}-\ell+1}  \right)$ and $\hatm \phi_\ell \triangleq  \left(\hat \phi_{\ell,1},\, \ldots,\, \hat \phi_{\ell,N_{\textrm{tx}}-\ell+1}  \right)$. With this, we can write $\m {\tilde V}_u$ following a Givens decomposition, so that the covariance matrix is
\begin{eqnarray}
 \m C_{u,j} &=& \field{E}\,\m {\tilde I}^*  \left(\prod_{\ell=1}^{L_u} \m D_l\left( \m \phi_\ell  \right) \prod_{n=\ell+1}^{N_{\textrm{tx}}} \m G_{n,\ell}\left(  \psi_{\ell,n} \right)\right)^*   \m F_k  \\ \nonumber && \m F_k^*\times \left(\prod_{\ell=1}^{L_u} \m D_l\left(  \m \phi_l  \right) \prod_{n=\ell+1}^{N_{\textrm{tx}}} \m G_{n,\ell}\left(  \psi_{\ell,n} \right)\right) \m {\tilde I}
\end{eqnarray}
where we parametrized the matrix random variable $\m {\tilde V}_u | \hatm V_{u}$ using the Givens parameters $\phi_{\ell,n} | \hat \phi_{\ell,n}$ and $\psi_{\ell,n} | \hat \psi_{\ell,n}$. If we assume that all the angles $\phi$ and $\psi$ are independent, then the expected value over all the angles can be decomposed into several expected values, each one over a different angle. This is a reasonable assumption for a MIMO channel with zero-mean Gaussian iid entires \cite{Ansari06}, for example. To simplify the computation of the covariance matrix, we work with a vectorized version of $\m C_{u,j}$, $\m c_{u,j} \triangleq \vec{ \m C_{u,j}}$.  Using properties of the Kronecker product \cite{henderson1981vec}, we have
\begin{equation}
\label{e:cvec}
\field E \m c_{k,i} =  \left(\hatm I^\T \otimes \hatm I^* \right)\left(  \prod_{\ell=1}^{L_u}  \m R_{\ell}^\T \prod_{n=\ell+1}^{N_{\textrm{tx}}} \m W_{\ell,n}^\T \right)^\textrm{T} \vec \left( \m F_j \m F_j^* \right)
\end{equation}
where 
\begin{equation}
 \label{e:Rl}
 \m R_\ell  \triangleq \field{E}_{\m \phi_\ell | \hatm \phi_\ell } \left\{ \m D_\ell^\T\left(\m \phi_\ell  \right) \otimes  \m D_\ell^*\left( \m \phi_\ell  \right) \right\}
\end{equation}
and
\begin{equation}
\label{e:W}
 \m W_{\ell,n} \triangleq \field{E}_{\psi_{\ell,n} | \hat \psi_{\ell,n}} \left\{ \m G_{n,\ell}^\T\left(  \psi_{\ell,n} \right) \otimes \m  G_{n,\ell}^*\left(  \psi_{\ell,n} \right)\right\}.
\end{equation}
We now proceed to approximate $ \m R_\ell$ and $ \m W_{\ell,n}$. We resort to a well known result in high resolution quantization theory \cite{Zamir94} and approximate the quantization error by a uniform random variable in the quantization bin. This approximation is exact for the quantization noise $\hat\phi_{\ell,i} - \phi_{\ell,i}$ (but not for  $\hat \psi_{\ell,n} -  \psi_{\ell,n}$)  if $\m H_{u}$ has independent and identically distributed Gaussian entries, since $\hat \phi_{\ell,i}$ is uniformly distributed in $[0,2\pi]$ \cite{Roh07}.

We now calculate a closed form expression for (\ref{e:Rl}) and (\ref{e:W}) assuming uniform quantization noise. For the sake of clarity, we will omit subscripts and matrix angular arguments when their value is clear. Using the definition in (\ref{e:Dl}), it is possible to take the Kronecker product and write (\ref{e:Rl}) as
\begin{equation}
\diag\left(\m D_\ell^*,\, \ldots , \, \m D_\ell^*, \,  e^{j\phi_{\ell,1}} \m D_\ell^*, \, \ldots \,  e^{j\phi_{\ell,N_{\textrm{tx}}-\ell+1}} \m D_\ell^*  \right).
\end{equation}
Now we compute the expectations of each term. First observe that
\begin{eqnarray}
\label{e:expD}
\field{E} [\m D_\ell^*]_{\ell+i-1,\ell+i-1} & =& \field{E} e^{-j\phi_{\ell,i}} = \frac{1}{2\epsilon}\int_{\hat \phi_{\ell,i} - \delta}^{\hat \phi_{\ell,i} + \delta} e^{-j\phi} \textrm{d}\phi 
\\ \nonumber &=& \frac{e^{-j\hat \phi_{\ell,i}}\sin \delta}{\delta}.
\end{eqnarray}
The expectation of the other diagonal terms can be obtained similarly as
\begin{equation}
\label{e:integration}
\field{E} e^{j\phi_{\ell,j}} [\m D_\ell^*]_{i,i} = \frac{e^{j\hat \phi_{\ell,j}}\sin \delta}{\delta},\, i=1\ldots \ell-1
\end{equation}
and
\begin{eqnarray}
\field{E} e^{j\phi_{\ell,j}} [\m D_\ell^*]_{\ell+i-1,\ell+i-1} =  \left\{ \begin{array}{l l}
                                                                     1 & \mbox{if } i=j\\
								     \displaystyle\frac{e^{j\left(\hat \phi_{\ell,j}- \hat \phi_{\ell,i} \right)}\sin^2\left( \delta \right)}{\delta^2} & \mbox{if } i \neq j
                                                                    \end{array} \right.
\end{eqnarray}
for $i = 1\ldots N_{\textrm{tx}}-\ell+1$. Now, we proceed to obtain a closed form for $(\ref{e:W})$ with the uniform error approximation. First, let us write $\m W_{\ell,n}$ as in (\ref{e:eWell}).

The expectations of the different blocks of the matrix can be obtained by simple integration, similarly to (\ref{e:integration}). The results for the different blocks are shown in (\ref{e:firstEq})-(\ref{e:lastEq}).

\begin{figure}
\begin{centering}
 \includegraphics[width=.9\figwidthuna]{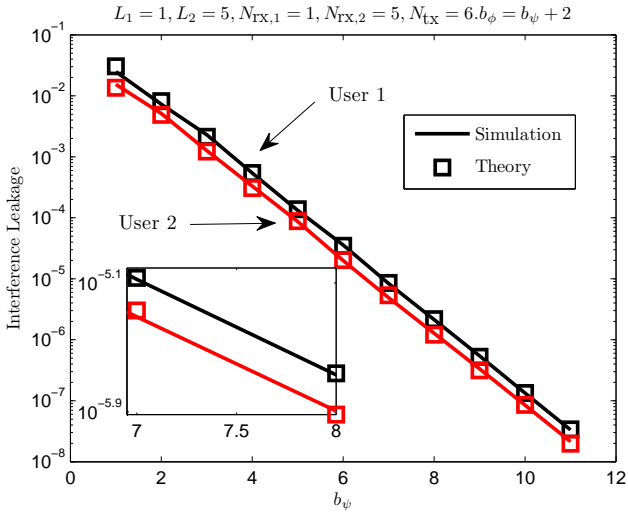}
\caption{Analytical and empirical interference leakage per spatial stream $\left(\frac{1}{L_u}\tr\mbox{ } \m C_{u,j}\right)$ for a two-user scenario. The theoretical approximation is calculated using (\ref{e:cvec}) and the closed form expressions for the expected values (\ref{e:expD}) - (\ref{e:finalExp}) . The simulation interference leakage was averaged over 100 independent MIMO channels with independent complex Gaussian entries.}
\label{f:approx}
\end{centering}
\end{figure}

Note that the complexity of obtaining the error covariance matrix is similar to the complexity of recovering the matrix $\hatm V$ from the quantized angles. Figure \ref{f:approx} shows the accuracy of the analytical approximation.

\section{MCS Selection}
\label{s:MCS}
The MCS selection block consists of a function $\mu$ that takes as input the set of post-processing SNR values of user $u$ and the number of spatial streams $L_u$, and computes the higher MCS that meets the FER constraint for those SNR values, i.e.
\setcounter{equation}{38}
\begin{equation}
\label{e:MCSsel}
 \mu\left(\m \gamma_u, L_u \right) = \arg \max_{c \in {\cal C } } \eta\left(c,L_u \right) \,\,\, \mbox{subject to   } \,\, \mathrm{FER }\left(\m \gamma_u, c_u \right) \leq p_0.
\end{equation}
The post-processing SNR values are calculated by applying the approximations (\ref{e:gu2}) and  (\ref{e:cvec}) in (\ref{e:postSNR})  to incorporate the interference leakage estimate. An important observation is that in (\ref{e:MCSsel}) the rate is being maximized, not the throughput. The reason is that for small values of $p_0$, the feasible points meet $\eta\left(c,L_u \right)\left(1-\mathrm{FER }\left(\m \gamma_u, c_u \right) \right) \approx \eta\left(c,L_u \right)$. This approximation simplifies the problem, since estimating the actual value of $\mathrm{FER }\left(\m \gamma_u, c_u \right)$ is not required, rather it is only necessary to discriminate whether it is above the desired threshold $p_0$ or not.

We use a machine learning inspired approach to solve (\ref{e:MCSsel}). Essentially, we classify features derived from the channel into the highest MCS that meets the target FER constraint. The classifier is made up of individual classifiers that distinguish whether a certain MCS and number of spatial streams are supported by the current channel. Note that this is slightly different from conventional machine learning in that there is a target average error rate, whereas machine learning usually involves avoiding classification errors altogether. 
We will follow a supervised learning approach to solve this problem, which includes two separated tasks: feature extraction and classification.

\subsection{Feature extraction}
In machine learning, the curse of dimensionality is well known: the larger the dimension of the feature vector, the exponentially more data is required \cite{Hughes68}. To reduce the dimensionality of $\m \gamma_u$, we exploit insights made in \cite{DanielsConv} about performance in coded bit interleaved MIMO-OFDM systems. In particular, it was recognized that performance was invariant to subcarrier ordering, i.e.,
$
\mathrm{FER }\left(\m \gamma_u, {c}_u \right) = \mathrm{FER }\left(\m \Pi\m \gamma_u, {c}_u \right)
$, 
with $\m \Pi$ any permutation matrix. Therefore, the reduced dimension feature vector should be invariant to subcarrier ordering. Similar to \cite{DanielsConv}, we use a subset of the ordered SNR values as our feature vector. Define the \textit{ordered SNR vector} $ \tilde {\m \gamma}_u = \left[ \tilde {\gamma}_{u,1}\,\ldots\, \tilde { \gamma}_{u,NL_u}\right]^T$ as a vector formed by ordering the elements in $\m \gamma_u$ in ascending order. For example, $\tilde { \gamma}_{u,i}$ denotes the $i$-th smallest SNR value (among all carriers and spatial streams) of user $u$. We obtain our feature vector $\m f$ by selecting a subset of the entries of $\tilde {\m \gamma}_u$. Other approaches for dimensionality reduction, like principal component analysis \cite{Wold198737}, may alternatively be applied, but in our simulations we did not see a significant benefit.

\subsection{Classification}
 The objective of the classification task is to estimate the highest MCS supported by the channel, as characterized by the feature vector $\m f$. Following a similar approach as in \cite{DanielsOnlineSVM}, we use a set of classifiers $\delta_{c, L}(\m f )$  to discriminate whether the current channel will support transmission with MCS $c$ and $L$ spatial streams while meeting the FER constraint. More formally, given a set of $M$ training samples $\left\{\left(\m \gamma_i,\, p_{\textrm{tr},i} \right) \right\}_{i=1}^M$, with $p_{\textrm{tr},i}$ the FER of the $i$-th channel, the input data for the classifier is the set $\left\{\left(\m f_i,\, \nu_i \right) \right\}_{i=1}^M$, with
\begin{equation}
\nu_i = \left\{ \begin{array}{ll}
1 & \textrm{if } p_{\textrm{tr},i} \leq p_0\\
-1 & \textrm{if } p_{\textrm{tr},i} > p_0 \\
\end{array}\right.
\end{equation}
and $\m f_i$ the subset of the ordered SNR of the $i$-th channel. The classifier is a function of the feature input vector that maps
\begin{equation}
\delta_{c,L}: \,\, \m f \rightarrow \left\{ -1, 1\right\}.
\end{equation}
There are several ways to construct a classifier; we choose the popular SVM. An SVM determines the class of a sample by the use of linear boundaries (hyperplanes) in high dimensional spaces. Operating in a high dimensional space is enabled by the use of a Kernel function $K(\m x_1, \m x_2)$ that maps $\m x_1$ and $\m x_2$ to vectors $\phi(\m x_1)$ and $\phi(\m x_2)$ lying in a Hilbert space, and performs the inner product in that space $\left\langle\phi(\m x_1),\phi(\m x_2) \right\rangle$. The Kernel function $K(\m x_1, \m x_2)$ has a very simple form for properly chosen $\phi$. In many cases, perfect separation by a hyperplane is not possible (or not desirable, since it would lead to non-smooth boundaries) and a penalization term is introduced to take into account the misclassified training samples. Formally, the classifier is
\begin{equation}
\delta_{c,L}(\m x) = \mbox{sign}\left( \sum_{i=1}^M \alpha_i \nu_i K\left(\m x, \m f_i \right) + b\right)
\end{equation}
where $\alpha_i$ is obtained as the result of the optimization problem
\begin{equation}
\begin{array}{ll}
\mbox{minimize }  & \frac{1}{2}\sum_{i=1}^M\sum_{j=1}^M \alpha_i \alpha_j K\left( \m f_i, \m f_j\right) - \sum_{i=1}^M \alpha_i \\
\mbox{subject to} & \sum_{i=1}^M \nu_i \alpha_i = 0 \\
		& 0 \leq \alpha_i \leq C,\, i=1,\, \ldots,\, M
\end{array}
\end{equation}
and $b$ can be obtained by solving  $\delta_{c,L}(\m f_i)\nu_i = 1$ for any training sample $\m f_i$ such that $0 < \alpha_i < C$ \cite{elementsStat}. The parameter $C$ has to be adjusted to trade off smoothness and training misclassification rate. High values of $C$ result in very irregular boundaries caused by very small training errors, and low values of $C$ result in large training errors caused by smooth boundaries.
In this paper we use the radial basis function kernel
\begin{equation}
K(\m x_1, \m x_2) = \exp \left(- \frac{\left\|\m x_1 - \m x_2 \right\|^2}{\rho^2} \right),
\end{equation}
where parameter $\rho^2$ is used to tradeoff bias and variance: small values of $\rho$ tend to take into account only the nearby training points, leading to high variance classifiers, and large values of $\rho$ result in biased results. The parameters $\rho$ and $C$ are selected using a cross-validation approach \cite{elementsStat}.

For a given number of streams $L_u$, the overall classifier chooses the MCS with a higher rate among those predicted to meet the FER constraint. The MCS selection function $\mu$ (\ref{e:MCSsel}) is implemented as
\begin{equation}
  \mu\left(\m \gamma_u, L_u \right) = \arg \max_{c \in {\cal C}} \left\{\eta\left(c, L_u \right) \right\} \,\, \mbox{s.t. } \,\, \delta_{c,L}(\m \gamma_u) = 1.
\end{equation}

\section{User and Mode Selection}
\label{s:userSel}
Performing optimal user and mode selection requires an exhaustive search over all possible combinations of users and number of streams per user. To overcome this challenge, we propose a greedy approach, similar to \cite{Dimic05, Kobayashi06, Chen08}, where the streams are added one by one until the utility function $\nu\left(\m t \right)$ does not increase. In each iteration, one spatial stream is added to the user whose increment in the number of spatial streams led to a higher throughput. The algorithm continues until the maximum number of spatial streams is reached, or when an increment in the number of spatial streams lead to a lower sum rate. 

An important observation is that the greedy algorithm is not fair, in the sense that a user could be assigned all the spatial streams if it led to a a higher objective function, in our case the sum rate. The objective function, however, could be modified to encourage fairness. For example, metrics that are concave in the rates would give more utility to assigining spatial streams to unscheduled users. This change in the objective function does not require any other change in the user selection algorithm, or the other pieces of our link adaptation procedure. The entire proposed link adaptation algorithm is summarized in Algorithm 1.

\begin{algorithm}
\begin{algorithmic}
\caption{Link Adaptation Algorithm}
 \STATE $L_u = 0 \,\ \forall u$
 \STATE ${\cal R} \leftarrow 0$
 \WHILE{$\sum_{u=1}^U L_u < N_\mathrm{tx}$}
    \FOR{Each user $u$ with $L_u < N_\mathrm{rx,u}$}
	\STATE Calculate matrices $\m F_v[n]$, $\m G_v[n]$, $\m B_v[n]$ for all users $v$, for the spatial streams set $\left\{ L_1, L_2,\, \ldots,\, L_u + 1,\, \ldots,\, L_K\right\}$ following the procedure in Section \ref{ss:precoding}.
	\STATE  Calculate interference leakage $\m R_{u,j}$ as in (\ref{e:approxInt}).
	 \STATE Calculate post-processing SNR values $\m \gamma_v \forall v$ as in (\ref{e:postSNR}).
         \STATE $c_v \leftarrow \mu\left(\gamma_v, L_v \right) \forall v$. \COMMENT{Calculate optimum MCS for all users}
	 \STATE $t_v \leftarrow \eta\left(c_v,L_v \right) \forall v$ \COMMENT{Calculate the corresponding rate}
	 \STATE ${\cal R}_u \leftarrow \nu\left( \m t \right)$ \COMMENT{Utility metric if we incremented $L_u$ by 1}
    \ENDFOR
    \STATE $j \leftarrow \arg \max_{u} \left\{ {\cal R}_u \right\}$ \COMMENT{User whose increment in $L_u$ leads to a higher rate}
    \IF{${\cal R}_j \geq {\cal R}$} 
	\STATE $L_j \leftarrow L_j + 1$
	\STATE ${\cal R} \leftarrow {\cal R}_j$
    \ELSE
	\STATE Stop algorithm.
    \ENDIF
 \ENDWHILE
\end{algorithmic}
\label{a:greedy}
\end{algorithm}

%
%
\section{Simulation Results}
\label{s:results}

\begin{table}
\centering
\caption{MCS in IEEE 802.11ac with the corresponding data rates in 20MHz channels \cite{11ac}.}
  \begin{tabular}{c  c  } \hline
MCS & Data Rate (Mb/s)\\ \hline \hline
BPSK 1/2 & 6.5 \\\hline
QPSK 1/2 & 13 \\\hline
QPSK 3/4 & 19.5 \\\hline
16-QAM 1/2 & 26 \\\hline
16-QAM 3/4 & 39 \\\hline
  \end{tabular}
 \begin{tabular}{c  c  c } \hline
MCS & Data Rate (Mb/s)\\ \hline\hline 
64-QAM 2/3 & 52 \\\hline
64-QAM 3/4 & 58.5 \\\hline
64-QAM 5/6 & 65 \\\hline
256-QAM 3/4 & 78\\\hline 
 & \\\hline
  \end{tabular}
\label{t:MCS}
\end{table}

To validate the performance of the proposed link adaptation method, we performed simulations using parameters from the physical layer of IEEE 802.11ac. The studied scenario comprises a 4-antenna transmitter communicating with three 2-antenna receivers over a 20MHz channel (52 OFDM carriers) with an 800ns guard interval. The frame length was set to 128 bytes. Perfect CSI was assumed at the receiver and different levels of CSI at the transmitter. The FER constraint for the link adaptation problem was set to $p_0 = 0.1$. The set of MCS for optimization with the associated data rate for one spatial stream is shown in Table \ref{t:MCS}.

The training of the classifiers $\delta_{c,L}$ was performed as follows. The training set was generated in a single user setting with perfect CSI by simulating different channels for all MCS and NSS values. For each of the channels, the complete transmit-receive chain was simulated, e.g. coding, interleaving, equalization, decoding. The resulting SNR $\m {\tilde\gamma}_i$ values and FER $p_{\textrm{tr},i}$ were stored for each of the channels. The channels were generated in the time domain with a 4-tap MIMO channel with iid Gaussian entries with 30 different noise levels, corresponding to SNR values between 5 and 50 dB. For each training sample, the feature vector $\m f$ consisted on 4 equispaced SNR values (including the first and last ones) from the ordered SNR vector $\m{\tilde \gamma}_i$. 
The class $\nu_i$ of each training sample was adjusted to $-1$ if the measured FER was above the desired threshold, i.e., $p_{\textrm{tr},i} > p_0$, and 1 otherwise. We do not consider multiuser or limited feedback CSI in the training set, since we assumed that the SNR information $\m {\tilde \gamma}_i$ was enough to predict the FER performance, regardless of whether the SNR values are the result of performing limited feedback BD precoding or not.
%

The parameters $\rho$ and $C$ of the SVM classifier were chosen before training the system. We followed the usual $K$-fold cross validation procedure \cite{elementsStat}  (with $K=4$) to select these parameters. The SVM was implemented with the \texttt{LIBSVM} software package \cite{CC01a}.

\begin{table*}[t]
\caption{Classification errors and accuracy gain of the SVM classifier with respect to the Average SNR classifier (all in \%).}
\scriptsize
\begin{tabular}{|l|l|l|l|l|l|l|l|l|l|l|l|l|l|l|l|l|l|l|l|l|}
\hline
& \multicolumn{4}{|c|}{SVM} & \multicolumn{4}{|c|}{Av. SNR} & \multicolumn{4}{|c|}{Gain Av. SNR} & \multicolumn{4}{|c|}{Eff. SNR} & \multicolumn{4}{|c|}{Gain Eff. SNR}\\ \hline &\textbf{$L$=1}&\textbf{$L$=2}&\textbf{$L$=3}&\textbf{$L$=4}&\textbf{$L$=1}&\textbf{$L$=2}&\textbf{$L$=3}&\textbf{$L$=4}&\textbf{$L$=1}&\textbf{$L$=2}&\textbf{$L$=3}&\textbf{$L$=4}&\textbf{$L$=1}&\textbf{$L$=2}&\textbf{$L$=3}&\textbf{$L$=4}&\textbf{$L$=1}&\textbf{$L$=2}&\textbf{$L$=3}&\textbf{$L$=4} \\ \hline
\textbf{BPSK 1/2}&0.45&1.33&1.05&0.88&9.4&2.9&1.78&2.82&95.2&54.0&41.1&68.6&0.866&1.2&1.08&1.46&48.1&-11.1&3.08&39.8\\\hline
\textbf{QPSK 1/2}&0.433&1.12&1&0.683&3.92&7.87&2.33&3.32&95.2&54.0&41.1&68.6&1.1&1.35&1.18&0.9&60.6&17.3&15.5&24.0\\\hline
\textbf{QPSK 3/4}&1.43&3.57&4.78&2.08&9.22&6.38&6.7&3.78&84.4&44.1&28.6&44.9&2.75&1.4&3.35&2.83&47.9&-154&-42.8&26.5\\\hline
\textbf{16-QAM 1/2}&0.417&0.6&1.22&0.8&9.73&4.11&1.97&2.45&95.7&85.4&38.1&67.3&1.51&1.58&1.68&1.8&72.5&62.0&27.7&55.6\\\hline
\textbf{16-QAM 3/4}&1.58&3.87&3.63&1.9&7.63&7.15&4.92&3.28&79.3&45.9&26.1&42.1&3.07&3.63&3.1&2.7&48.4&-6.42&-17.2&29.6\\\hline
\textbf{64 QAM 2/3}&0.433&1.5&1.66&1.15&4.5&5.11&3.31&2.36&90.3&70.6&49.7&51.4&1.72&2.52&2.33&2.87&74.8&40.4&28.6&59.9\\\hline
\textbf{64 QAM 3/4}&1.03&2.68&3.05&2.05&6.53&6.1&5.47&3.12&84.2&56.0&44.2&34.2&2.57&3.28&3.08&4.13&59.7&18.3&1.08&50.4\\\hline
\textbf{64-QAM 5/6}&0.783&1.85&1.57&2.33&7.87&6.38&2.82&5.67&90.0&71.0&44.3&58.8&2.57&3.03&1.88&3.01&69.5&39.0&16.8&22.6\\\hline
\textbf{Average}& \multicolumn{4}{|c|}{1.65} & \multicolumn{4}{|c|}{5.03} & \multicolumn{4}{|c|}{67.2} & \multicolumn{4}{|c|}{2.24} & \multicolumn{4}{|c|}{26.3}\\ \hline
\end{tabular}
\label{t:resul}
\end{table*}

We compare the performance of the SVM classifier with an average SNR and an exponential effective SNR classifier. The training and the test sets were generated independently, each one containing 6000 samples. For each sample, the FER was estimated by simulating the transmission of $10^3$ frames for each of the generated channels, so a small part of the classification errors may be caused by an imperfect FER estimation. This error, however, is expected to affect all the classifiers in the same way. The average SNR classifier discriminates the class using a threshold $\gamma_{\textrm{th}}$ on the average SNR, designed to minimize the training set error. Formally, the average SNR classifier is
\begin{equation}
\delta_{c,L}^{\textrm{Av. SNR}}\left( \m \gamma\right) = \mbox{sign} \left( \frac{1}{NL_u} \sum_{l=1}^{L_u}\sum_{n=1}^{N} \gamma_l\left[n\right] - \gamma_{\textrm{th}} \right).
\end{equation}
 The exponential effective SNR consists on a generalized mean of the SNR values
\begin{equation}
\gamma_{\textrm{eff}} = -\frac{1}{\beta}\log\left( \frac{1}{NL_u} \sum_{l=1}^{L_u}\sum_{n=1}^{N} \exp\left( - \beta \gamma_l\left[n\right] \right) \right)
\end{equation}
that is compared with a threshold $\gamma_{\textrm{eff, th}}$, designed to minimize the training set error. The parameter $\beta$ depends on the MCS, and is also selected to minimize the training set error. We can write the effective SNR classifier as
\begin{equation}
\delta_{c,L}^{\textrm{Eff. SNR}}\left( \m \gamma\right) = \mbox{sign} \left( \gamma_{\textrm{eff}} - \gamma_{\textrm{eff,th}} \right).
\end{equation}

 In Table \ref{t:resul} we show the accuracy results for the tested classifiers. We can see that the SVM classifier outperforms the average and effective SNR classifiers, and in many cases the classification error is below 1\%. For example, for 16-QAM 1/2, $L=1$, the SVM classifier misclassifies around 1 sample out of 200 (0.417\%), whereas the average SNR classifier makes approximately 20 times more errors (9.73\%). In that case, the error rate of the effective SNR classifier is 1.51\%, approximately 4 times more than the SVM classifier. The classification gain with respect to the Av. / Eff. SNR classifier  is calculated as $\left( \mbox{Error}_{\mbox{Av. / Eff. SNR}} - \mbox{Error}_{\mbox{SVM}} \right) / \mbox{Error}_{\mbox{Av. / Eff. SNR}}$, and in the 16-QAM 1/2 case for the Av. SNR classifier is approximately $\frac{20-1}{20} \approx 0.95$. This classification gain reflects the percentage of errors made by the average or effective SNR classifier that would be corrected by the SVM. For example, a classification gain of 100\% means that the SVM classifier is perfect, and a classification gain of 0\% means that the SVM has the same accuracy as the average SNR classifier. We can see that in some MCS the Eff. SNR classifier outperforms the SVM classifier, but in average the SVM performs 22.34\% better than the Eff. SNR classifier. The average gain with respect to the Av. SNR classifier is much higher, above 67\%. This result confirms that SVM-based link adaptation algorithms outperform state-of-the-art effective SNR classifiers, currently used widely in LTE, among other communication systems \cite{taotao11}.

\begin{figure*}[t]
\centering
\begin{subfigure}{.45\textwidth}
\begin{centering}
\includegraphics[width=\textwidth]{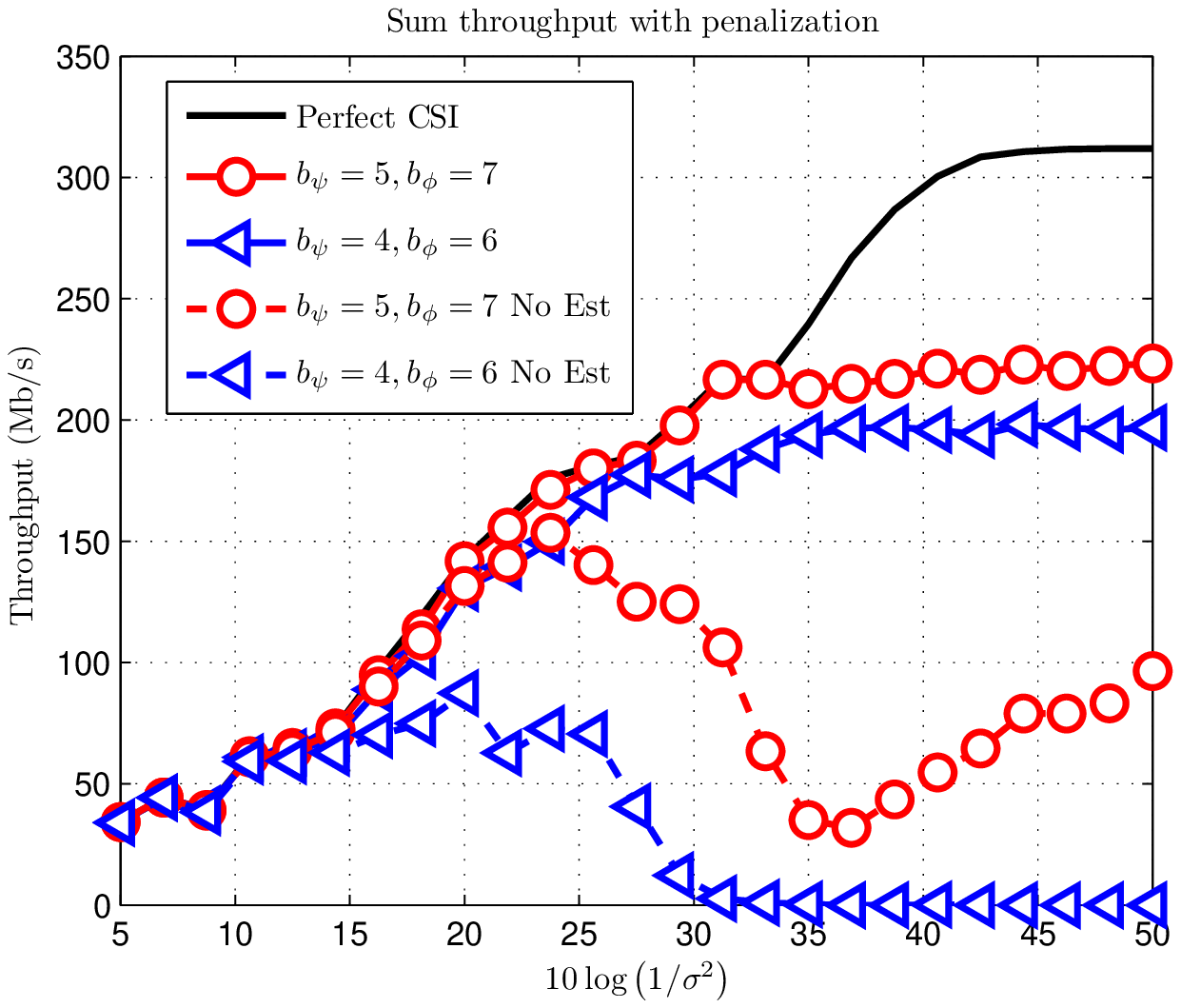}
\caption{Sum throughput.}
\label{f:resulTh}
\end{centering}
\end{subfigure}
\begin{subfigure}{.45\textwidth}
\centering
  \includegraphics[width=\textwidth]{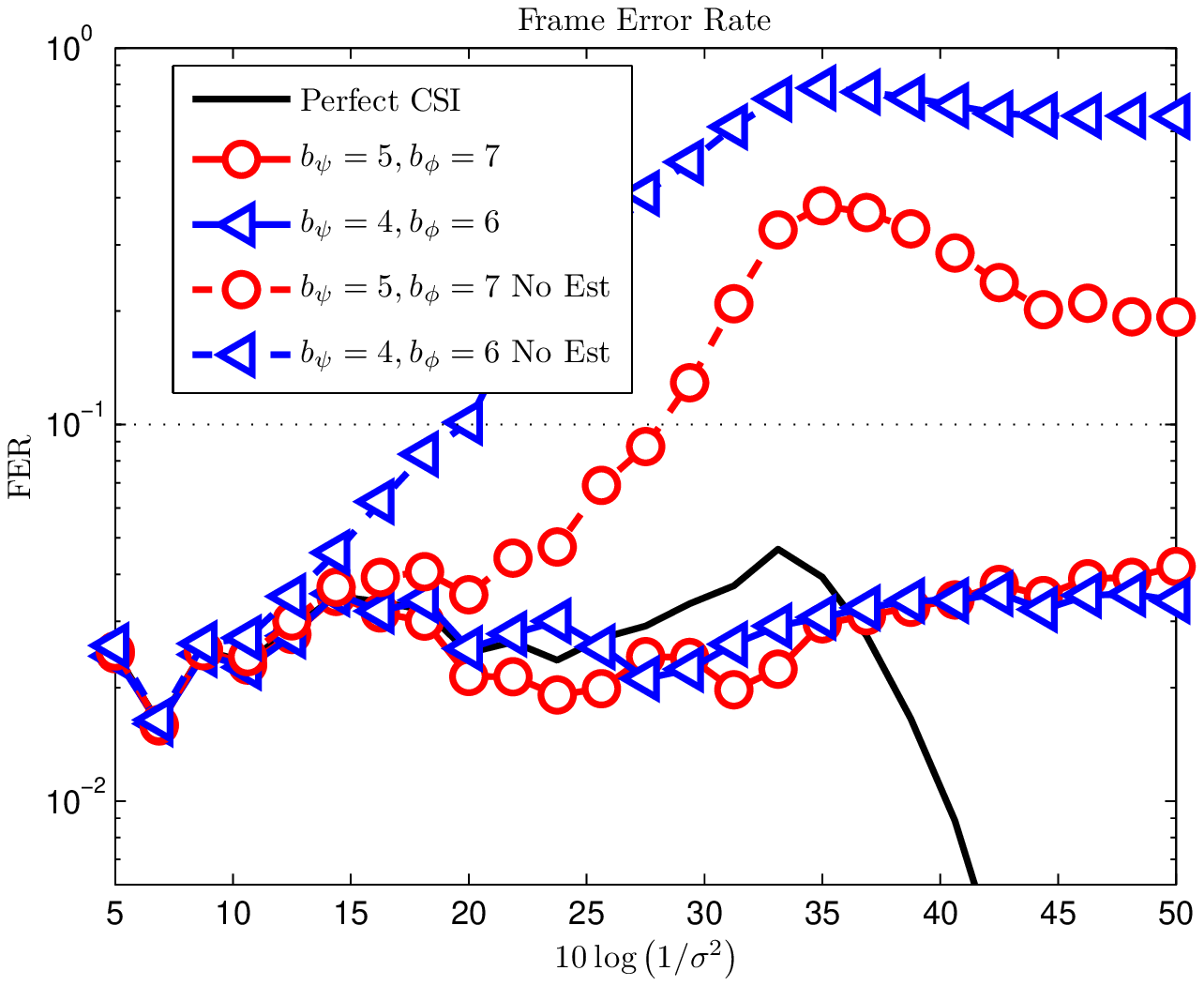}
\caption{Frame error rate.}
\label{f:resulFERR}
\end{subfigure}
\caption{Simulation results for different feedback rates, with and without interference estimation. 4-tap Gaussian channel model.}
\label{f:chanB}
\end{figure*}
\begin{figure*}[t]
\begin{centering}
 \includegraphics[width=1.9\figwidth]{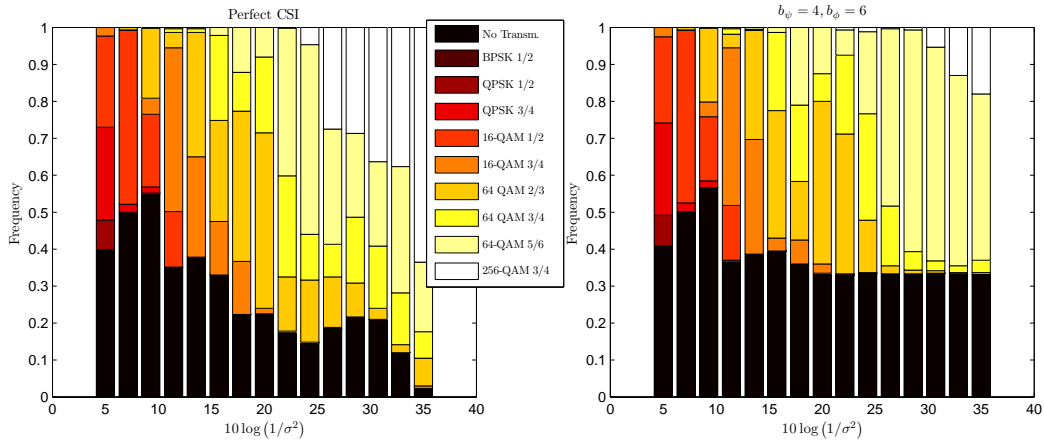}
\caption{Evolution of the frequency of the different MCS with the average SNR for the perfect CSI case and limited feedback with interference estimation.}
\label{f:MCS}
\end{centering}
\end{figure*}
\begin{figure*}[t]
\begin{centering}
 \includegraphics[width=1.9\figwidth]{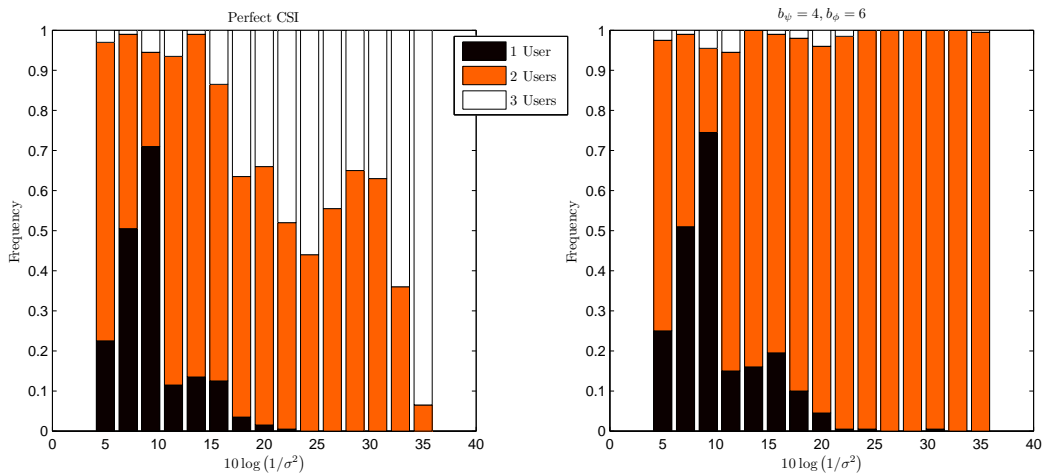}
\caption{Evolution of the frequency of the number of users scheduled for MU transmission with the average SNR for the perfect CSI case and limited feedback with interference estimation.}
\label{f:numUsers}
\end{centering}
\end{figure*}
\begin{figure*}[t]
\centering
\begin{subfigure}{.45\textwidth}
\begin{centering}
 \includegraphics[width=\textwidth]{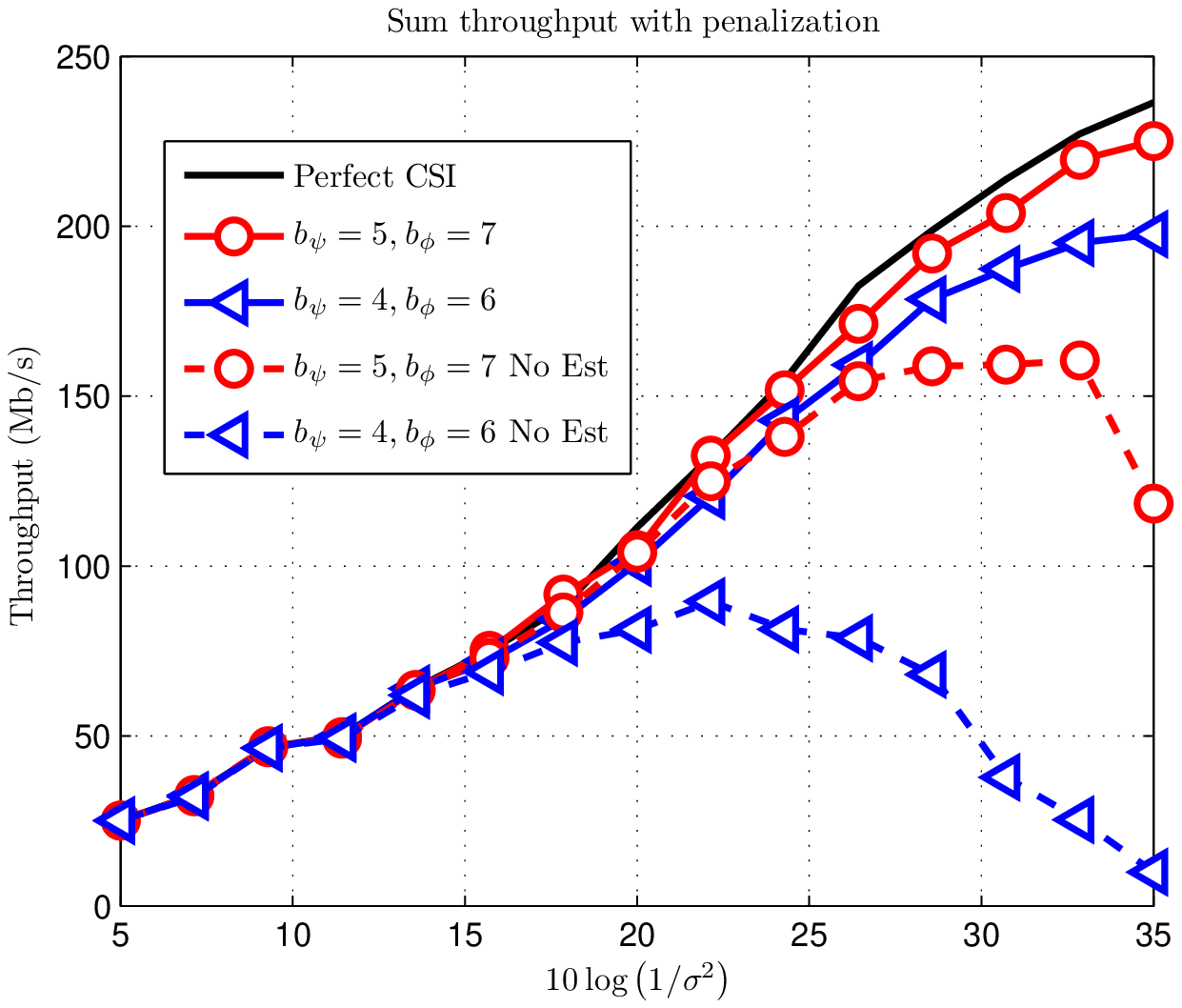}
\caption{Sum throughput}
\label{f:resulTh-chanB}
\end{centering}
\end{subfigure}
\begin{subfigure}{.45\textwidth}
\centering
 \includegraphics[width=\textwidth]{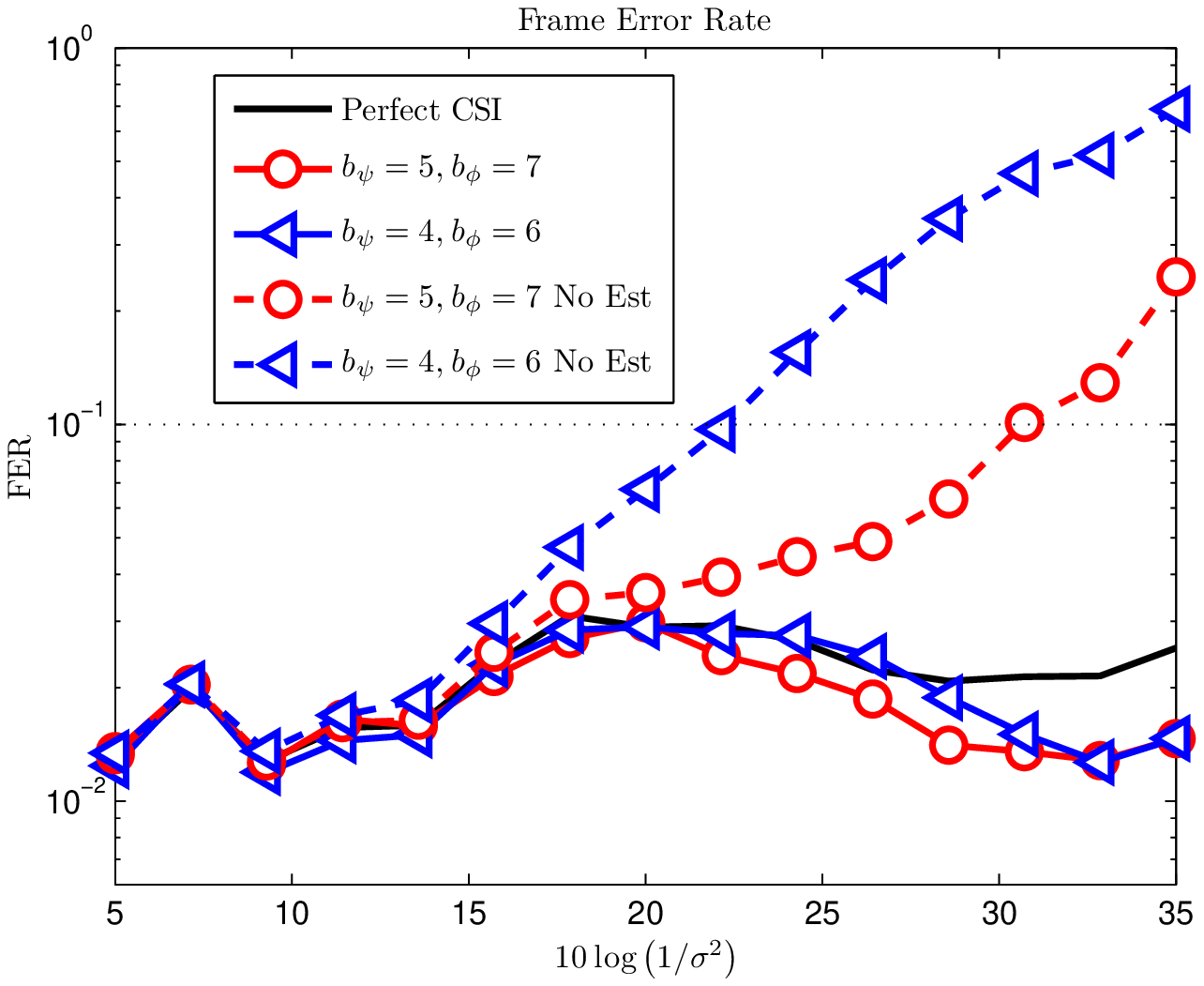}
\caption{Frame Error Rate }
\label{f:resulFERR-chanB}
\end{subfigure}
\caption{Simulation results for different feedback rates, with and without interference estimation. TGac channel model B. Classifier trained with Gaussian channel model.}
\label{f:chanB}
\end{figure*}

Complete system simulations were run for three different CSI levels at the transmitter: perfect CSI and limited feedback CSI with $\left( b_\psi = 5, b_\phi = 7 \right)$ and $\left( b_\psi = 4, b_\phi = 6 \right)$. Simulations were also run for the $\left(b_\psi = 6, b_\phi = 8\right)$ case, the higher rate available in IEEE 802.11ac, but the results are not shown as they are indistinguishable from the perfect CSI ones. In all cases a MU feedback was assumed (i.e., including SNR information of all carriers), even for $\left( b_\psi = 4, b_\phi = 6 \right)$, which is only available for SU mode in IEEE 802.11ac. The misclassified samples, i.e., the cases where the selected MCS led to a FER value over the predefined threshold, were penalized and computed as zero throughput. This penalization is set to remove the advantage of selecting an MCS with a higher throughput, but without meeting the outage constraint, and is a common procedure when evaluating link adaptation algorithms \cite{DanielsConv,Choi08}. The channels were generated independently from the training samples, but with the same statistical distribution. The channels for the three receivers were generated independently and with the same distribution, so fairness is automatically induced in this scenario. In other scenarios with users with different average received power, fairness can be introduced by changing the objective function, as explained in Section \ref{s:prEst}.

Figure \ref{f:resulTh} illustrates the sum throughput as a function of the noise variance. As expected, in the perfect CSI case, the throughput increases with the SNR, as the interuser interference is completely avoided in this setting, and lower noise levels enable the use of higher rate MCS. For the limited feedback CSI we study two different cases. The first case, represented by solid lines, is the result of following the complete link adaptation algorithm, including the interference estimation procedure. The second case, in dashed lines, does not include the interference estimation block (i.e., assumes $\m C_{u,j} = \m 0$), thus is overestimating the actual SNR. For the interference estimation case, we see that $\left( b_\psi = 5, b_\phi = 7 \right)$ follows the trend of the perfect CSI curve with a slightly lower throughput until an SNR value of around 35dB, and then flattens with a maximum throughput around 220Mb/s. Performance loss is more significant with the lower rate feedback $\left( b_\psi = 4, b_\phi = 6 \right)$ for moderate SNR values, reaching an error floor around 190Mb/s in the interference limited regime.  We can see that the perfect CSI curve flattens at 312Mb/s, which is the throughput of 4 streams using 256-QAM 3/4, the maximum rate MCS in Table \ref{t:MCS}. If the interference estimation is not performed (the curves tagged as \textit{No Est}) then the results are dramatically different. At low SNR values, the evolution of these curves is similar to the more sophisticated algorithm, as in this region the system is noise limited rather than interference limited. As the SNR increases, there is a huge degradation with respect to the perfect CSI case. Performance is penalized at high SNR where throughput decreases dramatically. This unexpected behavior is caused by the overestimation of the SNR due to the interuser interference that is not taken into account, which causes the classifier to choose an MCS with a higher rate than the channel can support. This leads to misclassified samples, leading to high FER values that drive the throughput towards zero. From a learning perspective, the feature set does not contain enough information to perform the adaptation. This information is implicitly included in the SNR values when performing interference estimation.

In Figure \ref{f:resulFERR}, we plot the evolution of the FER with the average SNR. We can see that in the perfect CSI case as well as in the cases where the interference is estimated, the FER constraint of $0.1$ is always met. This result shows the robustness of the proposed approach even with limited feedback. If the interference is not estimated, then the FER grows up to 1 due to the mismatch between the selected MCS and the actual SNR values, which causes the throughput to decrease, as previously explained. The classifier is able to correctly perform MCS selection in a multiuser setting despite being trained in a single user scenario, thus showing also the robustness of the classifiers against changes on the channel distribution. This plot shows the importance of performing the interference estimation procedure.

In Figure \ref{f:MCS}, we plot the fraction of the selected MCS for different SNR values, for both perfect CSI and limited feedback. In the perfect CSI case, the number of scheduled users is usually low for low SNR values (the \textit{No Transmission} frequency, representing the case where a user is not scheduled for transmission, is quite high) and the MCS are robust, while for higher SNR the \textit{No Transmission} frequency decreases and higher rate MCS are employed for transmission. In the limited feedback case the behavior is similar to perfect CSI for low SNR values, but for higher SNR the interuser interference forces to use more robust MCS and approximately one third of the time the \textit{No Transmission} MCS is selected.

Now we study the frequency of a given number of users being selected in Figure \ref{f:numUsers}. In the perfect CSI case the number of users grows up to three (i.e., all users are scheduled most of the time), while in the limited feedback case transmits to only two users most of the time. The reason is that in the limited feedback case the interuser interference increases with the scheduled number of users, which is the limiting factor in the high SNR regime. The user selection algorithm is able to identify the suboptimality of an aggressive multiuser transmission by estimating the residual interuser interference. The effect of scheduling only two users out of three was also observed in Figure \ref{f:MCS}, where for high SNR values the \textit{No transmission} MCS is selected 33\% of the time. This partial scheduling of the users can lead to unfair scenarios where the maximization of the sum rate causes one user not to be scheduled. This behavior can be corrected by changing the objective function of the LA problem, as explained in Section \ref{s:prEst}.

We also  run simulations using TGac channel model B \cite{ChanModWifi} to show the robustness of the proposed approach against changes in the environment or channel model. The classifier is the same as in the previous section, i.e., is trained with 4-tap MIMO channels with iid Gaussian entries. The 802.11 channel model B presents realistic characteristics not present in the 4-tap channel, such as delay taps with different power, and correlation among antennas. The objective of this experiment is to show if the ordered SNR feature vector suffices to characterize the performance of the system even in the presence of a change in the environment. The throughput and frame error rate results are shown in Figure \ref{f:chanB}. We can see that the link adaptation algorithm is able to keep the FER under the required threshold. Once again, interference estimation is shown to be necessary to correctly select the number of users and MCS.

\begin{figure}
\begin{centering}
 \includegraphics[width=.5\textwidth]{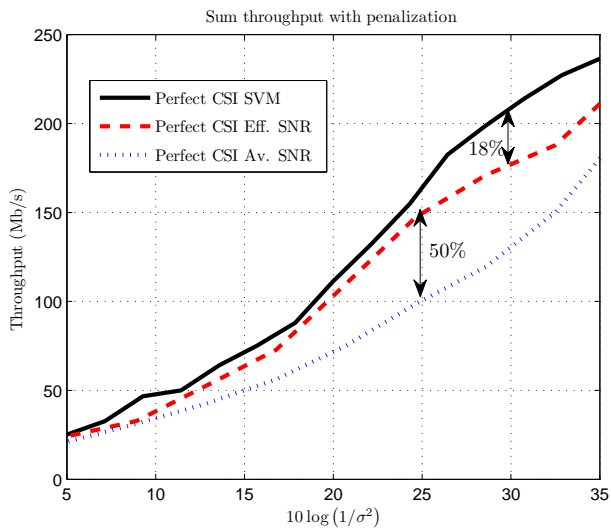}
\caption{Comparison between SVM, Eff. SNR and Av. SNR classifiers}
\label{f:comparison}
\end{centering}
\end{figure}

We compared the performance of the proposed adaptation procedure with the same algorithm, but using an effective SNR and average SNR classifier instead of SVM. We run the simulations under TGac channel model B, and assuming perfect CSI at the transmitter. The results are shown in Figure \ref{f:comparison}. The average SNR classifier offers a bad performance, as predicted by the results in Table \ref{t:resul}. The effective SNR has a slightly worse performance than SVM, especially at low SNR values. At moderate SNR values (around 20dB), SVM offers approximately a 20\% gain in throughput.

%
%
\section{Conclusion and Future Work}

In this paper, we presented a framework for link adaptation in multiuser MIMO-OFDM networks with limited feedback information.
The link adaptation problem is formulated as a maximization of the sum rate subject to a FER constraint. 
Focusing on the multiuser capabilities in IEEE 802.11ac, we developed a data-driven algorithm that performs user selection, mode selection, MCS selection and takes into account limited feedback information. 
We showed that estimating the interference due to imperfect CSI is crucial to achieve a good performance in the multiuser MIMO setting.
Depending on the feedback rate, and by estimating the residual interference, the transmitter is able to identify the error floor caused by multiuser transmission in the high SNR regime.
We conclude that machine learning classifiers can be used in a multiuser setting, even in limited feedback scenarios. Performing interference estimation, however, is crucial for the algorithm performance.
The machine learning classifier is also shown to be robust to changes in the statistical distribution of the channel. For example, the information acquired in an open environment can be used to effectively perform adaptation in an office setting, where more frequency selectivity is expected.

Future work includes incorporating more realistic system aspects. For example, we assumed a constant codeword length, whereas in IEEE 802.11ac the codeword length can vary from 1 to 65535 bytes, thus severely affecting the FER performance. It is not feasible to have a different classifier for every possible length, so a different approach has to be studied. With the proposed framework, a possible solution could be to include the packet length as an extra feature in the MCS classifier \cite{RicoHeathAsilomar13}.

Another limitation of the proposed framework is the assumption of ZF receivers. In a practical scenario, different receivers can implement different detection algorithms, and the link adaptation algorithm should be able to adapt to that as well. Although it is unlikely that the transmitter is able to know the exact algorithm the receiver is using, having a different MCS classifier for each receiver should be able to capture its effect (for example, the same SNR vector would lead to different FER values depending on the complexity of the detection algorithm). A similar limitation is the assumption of perfect CSI at the receivers. In practice, the receivers will estimate the channel based on the incoming signal, thus creating some CSI inaccuracy. This error is likely to affect the design of both the precoders and receive equalizers. Also, different receivers may use different channel estimation algorithms (e.g., a basic receiver is expected to use the pilot symbols to estimate the channel, whereas more advanced receivers may be able to improve channel estimation by exploiting the data symbols), and this behavior should be also captured by the learning algorithm.

We also assumed prior knowledge of a training set. In practice, the training samples have to be acquired online, thus suggesting an \textit{exploration versus exploitation} tradeoff. Finally, prototyping the system would require taking into account complexity and storage requirements.
\label{s:conc}

\bibliography{IEEEabrv,biblio}

\bibliographystyle{ieeetr}

\end{document}